\documentclass{svmult}

\usepackage[pdftex]{graphicx}
\usepackage[tight,footnotesize]{subfigure}
\usepackage{numprint}
\npthousandsep{,}

\begin{document}
\pagestyle{headings}  %
\tableofcontents
\mainmatter              %
\title{Partitioning Breaks Communities}
\titlerunning{Partitioning Breaks Communities}  %
\author{Fergal Reid \and Aaron McDaid \and Neil Hurley }
\authorrunning{Reid, McDaid and Hurley} %
\tocauthor{Fergal Reid, Aaron McDaid, Neil Hurley}
\institute{Clique Research Cluster, \\Complex Adaptive Systems Laboratory,\\ University College Dublin, Ireland\\
\email{fergal.reid@gmail.com}}

\maketitle              %

\begin{abstract}
\\Considering a clique as a conservative definition of community structure, we examine how \textit{graph partitioning algorithms} interact with cliques. Many popular community-finding algorithms partition the entire graph into non-overlapping communities.
We show that on a wide range of empirical networks, from different domains, significant numbers of cliques are split across the separate partitions produced by these algorithms.
We then examine the largest connected component of the subgraph formed by retaining only edges in cliques, and apply partitioning strategies that explicitly minimise the number of cliques split.
We further examine several modern overlapping community finding algorithms, in terms of the interaction between cliques and the communities they find, and in terms of the global overlap of the sets of communities they find.
We conclude that, due to the connectedness of many networks, any community finding algorithm that produces partitions must fail to find at least some significant structures.
Moreover, contrary to traditional intuition, in some empirical networks, strong ties and cliques frequently do cross community boundaries; much community structure is fundamentally overlapping and unpartitionable in nature.
\end{abstract}

\section{Introduction}
Groups of interacting entities can be considered as a complex system.
It is popular to examine such systems in terms of the networks their component entities form, to gain insight into properties of the system as a whole.
For example, the speed with which a contagion can spread through a system is partly determined by the topology of its underlying network.
The way subgroups of entities interconnect is also important to investigate whether useful higher level abstractions -- above the level of individual entities -- exist in the systems we study.
To find such structures, an extensive variety of algorithms have been developed, which attempt to find groups of nodes in the network that are structurally significant in some way; these groups are referred to in the literature as `communities'.
See Fortunato \cite{fortunato2009community} for an extensive review of these algorithms, which we will refer to as \textit{Community Finding Algorithms}, or \textit{CFAs}.

CFAs have been put to a range of applications, across several domains.
As CFAs are applied ever more broadly, it is important that the structures they find, and the consequences of the design choices that define them are well understood.
Particular CFAs should not be assumed to work across all complex networks, merely because they have evaluated well on some.
In this research, we argue that certain algorithms, notably CFAs that produce \textit{partitions} of the original network, return incomplete lists of the significant community structure, for at least some empirical networks.
We perform an in-depth analysis of how several different CFAs interact with the cliques present in empirical networks, and discuss the consequences of this analysis for our intuition about community structure.
We show that certain networks do not lend themselves well to partitioning, and caution against using partitioning algorithms as universal community finding tools.
\subsection{Cliques as Lower Bound Communities}
Each CFA finds structure that corresponds to a particular intuition of what a `community' is; however there is little agreement on how exactly to define community.
One common idea is that a community should have a high density of edges among its nodes, where \emph{density} refers to the ratio of the number of actual edges between the nodes in the community to the maximum possible number of edges between these nodes.
The bound of this definition is the graph theoretic structure known as a `clique' -- a fully connected subgraph, in which each node is connected to every other.
Cliques, as discussed by Luce and Perry \cite{luce1949method}, have long been considered as community structure in human social networks.
In the domain of social networks, this is particularly intuitive; if a user is friends with several others on Facebook, all of whom are also friends, then this is a significant structure of common friends.
In addition to this intuitive appeal, cliques are rare structures in the networks we study; due to the strict requirement for each node to connect to every other, clique structure is unlikely to arise by chance in a sparse network.
Cliques are thus important structures.
However, to define communities solely as cliques is very strict and conservative, for if even one connection in the group is missing -- perhaps due to an incompletely observed network -- then the found community will shrink.
Many CFAs thus try to find communities comprised of groups of nodes which are highly connected, but less connected than perfect cliques. However, we posit that a clique is a good conservative \textit{lower bound estimate} of community structure, in so far as an observed clique more than likely is wholly contained inside \emph{some} real-world community. A maximally interconnected group of nodes, in a sparse network, always represents an interesting structure.
\subsection{Partitioning Community Finding Algorithms}
Many leading CFAs assign communities by partitioning the network, that is, grouping the nodes into \emph{disjoint} subsets, assigning each node to exactly one subset.
This partitioning approach to community finding has become popular, perhaps due to the appeal of treating a complex network as a graph, and the body of literature on graph partitioning problems.
Early applications of graph partitioning, such as the applications of the Kernighan-Lin algorithm \cite{kernighan1970efficient} discuss problems that explicitly require partitions, such as electronic component layout.
However, in this work we are concerned about the completeness of the lists of community structure found by algorithms when used in other domains, such as social networks, and in complex networks generally.
Regarding cliques as underestimates of community structure, we believe that regardless of what specific structures a given CFA finds, to be thorough, it should find, for each clique, \textit{at least} one structure which is a superset of that clique.
A CFA -- considered as a tool that reveals structure in a complex network -- that returns no community in which a group of fully connected nodes are assigned together, is neglecting to provide a complete list of the structures in the network.
This is especially true if the clique is large in size.  %
\subsection{Related Work}
We show that in many complex networks, partitioning CFAs split cliques occurring within the network; and hence fail to find complete lists of the network structure.
We examine why this occurs, investigating the intuition underlying many partitioning CFAs, and their relationship with cliques.
We show, using cliques as a tool, that some traditional intuition describing communities as well connected sets of nodes, separated by narrow bridges, is not always correct.
Instead, many of the graphs we study exhibit a structure that can be better explained as the `pervasive overlap' discussed in \cite{ahn2010link}, \cite{evans2009line} than as independent, weakly-connected modules.
We analyse cliques, rather than any other community structure proposed in the `overlapping community finding' literature, because we require a definition of structure that is a fundamental, conservative, and convincing underestimate of community; for every community, we want to find a conservative subset of that community.
We use cliques, rather than structures such as the percolated $k$-cliques of Palla et al. \cite{palla2005uncovering}, because with percolated $k$-cliques, we find no universal $k$ consistent across networks with which to evaluate partitioning; this would make it difficult to be conservative in our analysis.
Rather than choosing a new definition of community and discussing whether it is sufficiently conservative, we instead use the fundamental definition of the clique and examine its implications in detail.
We analyse some of the same data as Leskovec et al. \cite{leskovec2008statistical}. However, while that influential work sought to investigate the quality of the best community structure, at each scale, by evaluating it in terms of \textit{conductance}, which penalises communities in proportion to their external edges, we instead investigate network structure from a different angle, by using the sociologically grounded idea of the clique to conservatively estimate community \emph{cores}.
We characterise to what level each and every clique is preserved after the network is partitioned, thus considering structures globally across the network. %

\begin{figure}[!htb]
\centering
\includegraphics[width=3.0in]{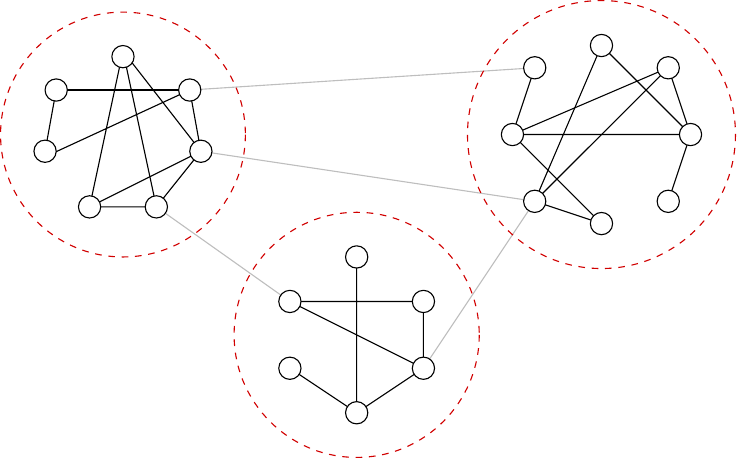}
\caption{Motivating image of network community structure from Newman \cite{newman2004detecting}}
\label{networkIntuition}
\vspace{-15px}
\end{figure}
An illustration of the intuition behind many CFAs can be seen in Figure \ref{networkIntuition}, from the influential paper by Newman \cite{newman2004detecting}, which shows separate and well-defined modules, connected by only narrow bridges.
This same intuition, conceptualising communities as connected by narrow bridges, can be traced back to the seminal work of Granovetter \cite{granovetter1973strength}:
\textit{``If the motivation to spread the rumor is dampened a bit on each wave of retelling, then the rumor moving through strong ties is much more likely to be limited to a few cliques than that going via weak ones; bridges will not be crossed.''}

Here, Granovetter is using `clique' in the sociological sense, closer to the modern idea of community, and the idea is that bridges -- narrow connecting links -- need to be crossed to carry information between such cliques.
This idea is further summed up in the modern review of Fortunato \cite{fortunato2009community} as: \textit{``If it were possible for a clique to move on a graph, in some way, it would probably get trapped inside its original community, as it could not cross the bottleneck formed by the inter-community edges.''}
However, this work, in keeping with research \cite{shi2007networks,van2006strong} on a limited number of other networks, finds evidence that structurally weak ties need not be crossed to traverse the network, contrary to the intuition just described.
In fact, we show that while the traditional intuition may be appropriate in some cases, the structure of many empirical networks does indeed lead to cliques crossing the `bottleneck' formed by inter-community edges.

\section{Experiments}

We conducted experiments to investigate the extent to which commonly used partitioning methods split the cliques in empirical network datasets.
To keep the number of cliques we consider tractable, and in keeping with the original sociological definition of clique \cite{luce1949method}, we constrain our analysis to \emph{maximal} cliques, which are cliques fully contained within no larger clique.
For convenience, we refer to maximal cliques as simply `cliques' in this work.
In our analysis, we first generate the complete list of cliques present in each network using the fast Bron Kerbosch algorithm \cite{bron1973finding}.
We then use the partitioning method under evaluation to assign each node to a community, and characterise how the cliques interact with the partitions found. We examine each maximal clique in turn, checking whether it is fully contained within a partition, or to what extent it has been split across partitions.
We quantify and present this metric for each network, initially using two distinct partitioning methods; one popular and efficient modularity optimization method \cite{blondel2008fast} and one normalised min-cut optimizing method \cite{dhillon2007weighted}.

\subsection{Network Datasets Examined}
To analyse data from a wide variety of networks, we gathered data from several different sources.
We used several network datasets from the SNAP project\footnote{http://snap.stanford.edu/data/} \cite{leskovec2008statistical}.
We examined networks formed by patterns of communication: The Enron and EU E-mail networks, and mobile telecoms data provided by an industrial partner\footnote{Idiro Technologies}, comprised of the voice call and SMS interactions on a mobile telecoms operator.
We examined relation networks formed in online social networks, consisting of several Facebook university network datasets \cite{traud2011comparing}, samples taken from the full Twitter follower network \cite{kwak2010twitter}, and the Slashdot online network.
For both Twitter and the Mobile telecoms data, where we had access to very large networks, beyond reasonable computational means to analyse, we generated 3 random snowball samples of each network to produce tractable datasets.
For the Facebook datasets, we chose to run our experiments on the smaller networks, due to the computational cost of calculating all maximal cliques.
We also considered the SNAP academic publication networks, the Web networks of Stanford and NotreDame, product recommendation networks from Epinions and Amazon, and Wikipedia voting network.
Finally, we considered a Protein-Protein interaction (PPI) network \cite{collins2007toward}, as an example of a biological network.

\subsection{Partition by Modularity Maximisation}
\begin{figure}[!htb]
\centering
\includegraphics[height=1.6in]{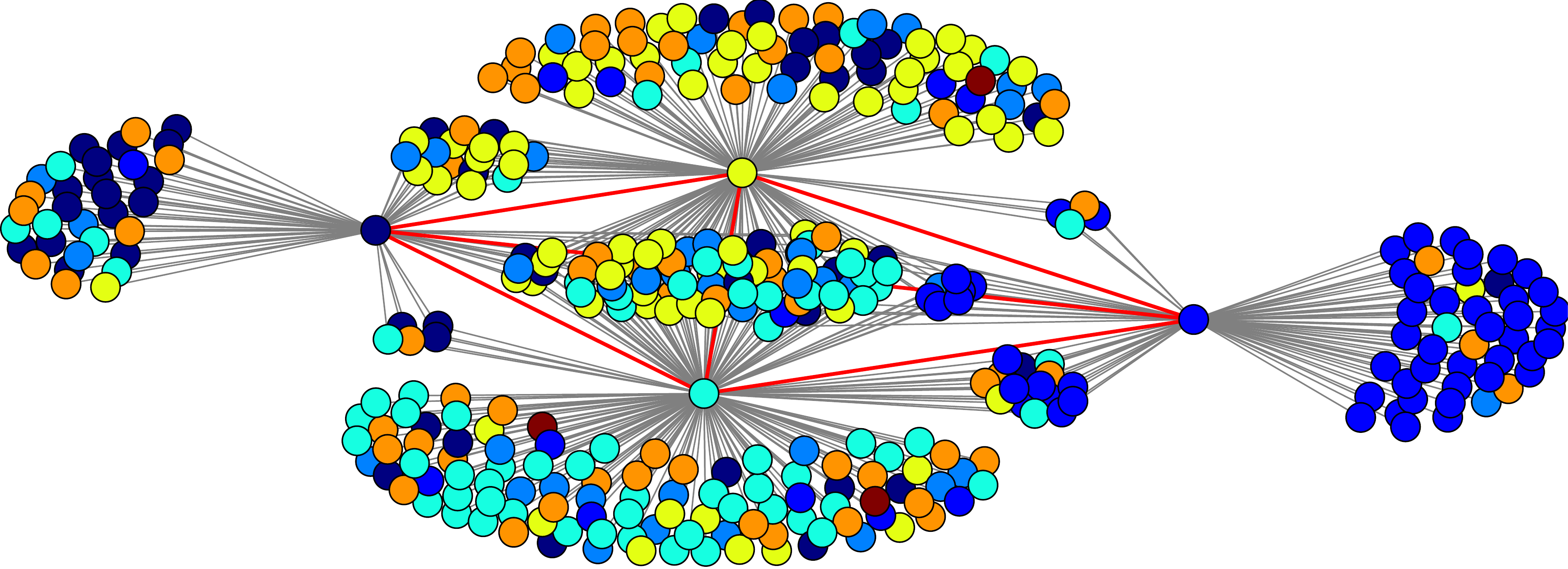}
\caption{Visualisation of one of the split 4-cliques from the Caltech Facebook dataset. Clique edges are shown in red; modularity partitions, as found by the Louvain method, are shown by color; as can be seen, each node of this 4-clique has been assigned to a different community. This clique will thus not show up in the list of found communities. Note the many paths of length 2 between the clique's nodes. }
\label{caltechSplitCliqueContainer}
\end{figure}

Many of the most popular CFAs are based on the modularity maximization approach of Newman \cite{newman2004detecting}.%
The modularity function measures community quality as a count of internal edges, less the expected number in a random graph with the same node degrees.
Modularity maximization algorithms, such as the fast `Louvain' method of Blondel et al. \cite{blondel2008fast} --  which we evaluate here -- designed to have a low computational cost on sparse graphs, and scale to large mobile call networks -- optimise for the number of partitions as well as the associated partitioning.
While traditional intuition holds that even triangles, or `strong ties', should not cross community boundaries, we are interested in more significant cliques -- so we initially restrict our analysis to cliques of size at least 4.
We also use a conservative definition of when a clique is `split' -- we say a clique is ``split at level $\alpha$'' if no partition contains more than $(100\times \alpha)\%$ of its nodes.
We quantify the proportion of cliques that are split by the partitioning of each network in two ways.
First, we examine the proportion of cliques of size at least 4 that are split at level $\alpha=0.9$.
Table \ref{damagedone} shows the significant proportions of cliques split at this level.
We would have expected, based on the traditional intuition, that such structures would be contained in the center of the found communities -- not spanning them, and not split by partitions that define found communities.
Figure \ref{caltechSplitCliqueContainer} provides an example of this effect, showing a single $4$-clique that has been split across 4 separate partitions by the community finding algorithm.

As our metric is the proportion of maximal cliques that have been split, we might be concerned that many of the maximal cliques will be small, such as 4-cliques, and that if a 4-clique is split by partitions -- while contrary to the intuition of structurally strong ties being unable to cross community boundaries -- this might not be of particular concern.
For a more conservative experiment, we consider only large cliques of size at least 8, split at level $\alpha=0.8$.
These parameters are arbitrary and we do not seek to justify them other than to reiterate that we are considering conservative structure, which would traditionally be expected only in the `cores' of found communities, not on their boundaries -- structure that a comprehensive CFA should return.
Even with this conservative definition, the partitions break significant numbers of such structures, on many networks -- see Table \ref{damagedone}.
For example, this shows that the Louvain CFA, run on the Caltech Facebook network, will split over one quarter of cliques of size 8 or more.%

\begin{table}[!ht]
\begin{center}
\caption{Proportion of maximal cliques split by the Louvain CFA, per network. We show the proportion of maximal cliques, of size 4 or greater, that have more than 10\% of their nodes assigned to different partitions (i.e.\ are split at level $\alpha$=.9). `Prop large cliques split' is the proportion of maximal cliques, of size 8 or greater, that that have more than 20\% of their nodes assigned to different partitions (i.e.\ are split at level $\alpha$=.8). `Cliques' is the number of maximal cliques in the network, `Partitions' is the number of partitions made by the Louvain method, and `Largest Clique' is the size of the largest clique in the network.}
\label{damagedone}
\begin{tabular}{llp{1.5cm}p{1.5cm}p{1.5cm}p{1.5cm}p{1.0cm}}
  \hline
  Network & Nodes & Partitions (As found by Louvain Method) & Maximal Cliques & Largest \par Clique & Prop. \par Cliques split & Prop. Large \par cliques split \\
    \hline
    \hline
    Email-Enron  & \numprint{36692} & \numprint{1363} & \numprint{205712} & 20 & 0.61 & 0.47 \\
    Email-EuAll  & \numprint{265009} & \numprint{15743} & \numprint{93267} & 16 & 0.82 & 0.67 \\
    Mobile1 & \numprint{10001} & \numprint{182} & \numprint{1550} &10 &0.97 & 0.00 \\  %
    Mobile2 & \numprint{10001} & \numprint{124} & \numprint{3538} &10 &0.90 & 0.00 \\  %
    Mobile3 & \numprint{10001} & \numprint{86} & \numprint{951} & 9&0.88 & 0.00\\ %
    \hline
    Facebook-caltech  & \numprint{769} & \numprint{10} & \numprint{31745} & 20& 0.68 & 0.27 \\
    Facebook-princeton  & \numprint{6596} & \numprint{21} & \numprint{1286678} & 34&0.44 & 0.22 \\
    Facebook-georgetown  & \numprint{9414}  & \numprint{26}  &\numprint{1440853}  & 33&0.41  &0.22  \\
    Twitter1  & \numprint{2001} & \numprint{8} & \numprint{23570} &12& 0.99 & 0.66 \\ %
    Twitter2  & \numprint{2001} & \numprint{4} & \numprint{554489} &27& 0.15 & 0.01 \\ %
    Twitter3  & \numprint{2001} & \numprint{7} & \numprint{130399} &22& 0.06 & 0.00 \\ %
    Slashdot0811  & \numprint{77360} & \numprint{771} & \numprint{441941} & 26&0.13 & 0.01 \\
    \hline
    Collab-AstroPhysics  & \numprint{18771} & \numprint{331} & \numprint{27997} & 57&0.60 & 0.32 \\
    Collab-CondMat  & \numprint{23133} & \numprint{626} & \numprint{8824} & 26&0.42 & 0.15 \\
    Collab-HighEnergy  & \numprint{9875} & \numprint{483} & \numprint{2636} & 32&0.23 & 0.00 \\
    Cite-HighEnergy  & \numprint{27769} & \numprint{172} & \numprint{419942} & 23&0.30 & 0.06 \\
    \hline
    Amazon0302  & \numprint{262111} & \numprint{173} & \numprint{117054} & 7&0.01 & 0.00 \\
    Epinions  & \numprint{75879} & \numprint{1607} & \numprint{1596598} & 23&0.38 & 0.11 \\
    \hline
    Web-NotreDame  & \numprint{325729} & \numprint{693} & \numprint{130965} & 155&0.04 & 0.00 \\
    Web-Stanford  & \numprint{281903} & \numprint{1013} & \numprint{774555} & 61&0.04 & 0.01 \\
    Wiki-Vote  & \numprint{7115} & \numprint{30} & \numprint{436629} & 17&0.65 & 0.37 \\
    \hline
    Protein-Collins & \numprint{1622} & \numprint{212} & \numprint{4310} &33& 0.16 & 0.08 \\
    \hline
    \end{tabular}
    \end{center}
    \end{table}

Our results show that the proportion of cliques split varies across the networks.
There is also a large variation in the number of maximal cliques present.
We might reason that this is due to some fundamental difference in the nature of the networks being considered, and question whether such analysis can be meaningfully applied across a range of networks.
After all, the Amazon network is a network of frequently co-purchased products, and the web datasets are explicitly constructed lists of hyperlinks; still other networks involve human communication or collaboration.
These networks are, however, frequently treated together as \textit{complex networks}; we might \textit{a priori} expect the same CFAs to perform well across them, and assume that a CFA proven in one domain will be automatically suitable and work well in other domains.
However, this modularity method seems to do poorly on some types of network, at least where finding complete lists of community is desired.
Similar results hold if we consider just the proportion of $n$-cliques split; as discussed in Section \ref{detailsection}.

\subsection{Relation of Modularity Found to Proportion Split}
\begin{figure}[!htb]
\centering
\includegraphics[width=2.9in]{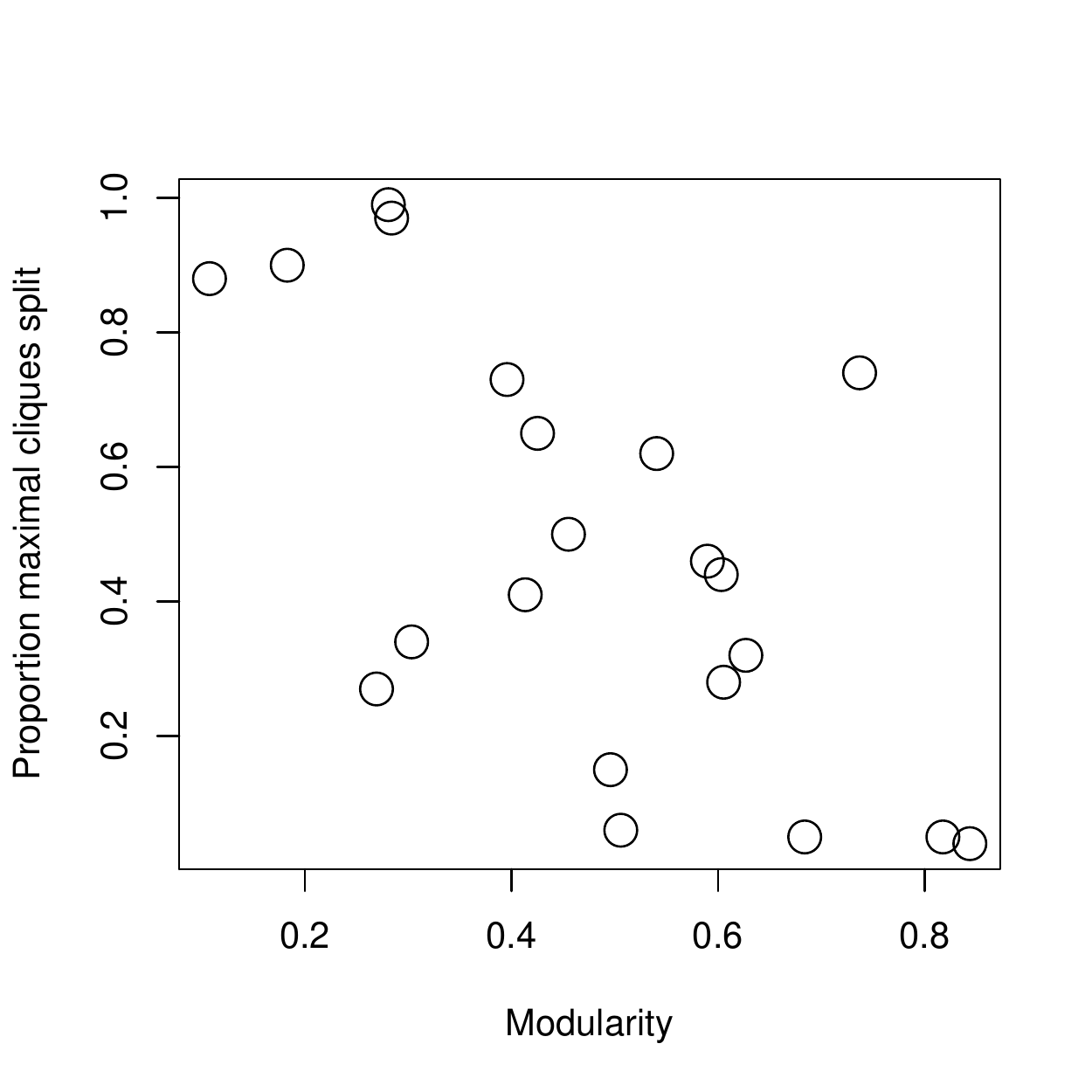}
\caption{Scatter plot of modularity of the partition vs the proportion of maximal cliques $>$10\% split (i.e.\ $\alpha$ =.9), for each network.}
\label{modularityDamageScatterplot}
\end{figure}

To investigate if the proportion of split cliques is in some way an artefact of low inherent modularity within the networks, we create a scatter-plot of the modularity achieved, against the proportion of maximal cliques split.
From Figure \ref{modularityDamageScatterplot} no obvious relationship appears.
Several of the network partitions have high modularity and still display significant clique splitting; if there is a fundamental characteristic that renders particular networks unsuitable for modularity based partitioning, in terms of the proportion of cliques that will be split, then the modularity achieved does not capture it.

\subsection{Partition by Normalized Edge Cut}
Another method that has previously been used for the purpose of community finding, from a different family of algorithms, is the multilevel kernel $k$-means partitioning method implemented in \textit{Graclus} \cite{dhillon2007weighted}, that minimises a normalized min-cut objective.
Like the modularity maximisation method of Blondel et al. this implementation is designed to scale to large networks, performing well on sparse data by avoiding expensive eigenvector computation.

We examined this method on the same data as the modularity maximisation method.
Unlike the modularity method, which discovers the number of partitions into which to break a graph, \textit{Graclus} requires this to be specified.
All other things equal, we would expect a smaller number of partitions would result in a smaller proportion of the maximal cliques being broken, and this effect is visible.
However, even when asked to produce a relatively small number of partitions -- relative to the network sizes -- min-cut partitioning results in large proportions of the cliques greater than size 4 being split on many datasets, as shown in Table \ref{damagedoneGraclus}.

\begin{table*}[!ht]
\caption{Proportion of cliques of size at least 4, split more than 10\% (i.e.\ $\alpha$=.9), by \textit{Graclus} \cite{dhillon2007weighted}, and \textit{hMETIS} \cite{karypis1999multilevel}, per network. Values shown for 4, 16 and 64 Partitions, with \textit{ufactor} 50, and 16 Partitions with \textit{ufactor}500. Also shown, proportion of the large connected component preserved, for subgraphs of edges in at least 4-Cliques `4-Clique', and edges in at least 5-Cliques `5-Clique'.}
\label{damagedoneGraclus}
\begin{center}
\begin{tabular}{l|p{0.8cm}p{0.8cm}p{0.8cm}|p{0.8cm}p{0.8cm}p{0.8cm}p{0.8cm}|lll}
\hline
Network & Graclus 4 & \ \par 16 & \ \par64 & hMETIS  4&\ \par16&\ \par64 &16 \par \textit{uf}500 & 4-Clique&5-Clique\\
\hline
\hline
Email-Enron  & 0.38 & 0.74 & 0.92 & 0.10 &0.54 & 0.67& 0.38&0.55 & 0.39\\
Email-EuAll  & 0.53 & 0.86 & 0.98 & 0.20 & 0.58 & 0.76 & 0.42&0.04 & 0.02\\
Mobile1  & 0.75 & 0.88 &0.99 & 0.47 & 0.81 & 0.93 &0.80 &0.17 & 0.07\\
Mobile2 & 0.66 & 0.93 &0.97 & 0.47 & 0.77 & 0.92& 0.64&0.20 & 0.09\\
Mobile3  & 0.83 & 0.93 & 0.96 & 0.48 & 0.82 & 0.95 &0.77 &0.06 & 0.01\\
\hline
Facebook-caltech  & 0.62 & 0.86 &1.00 & 0.56 & 0.89 & 0.99& 0.57&0.89 & 0.84\\
Facebook-princeton  & 0.33 & 0.69 &0.89& 0.32 & 0.58 & 0.89 &0.36&0.92 & 0.89\\
Facebook-georgetown  & 0.30 & 0.58 &0.80 & 0.32 & 0.50 & 0.74 &0.40&0.93 & 0.90\\
Twitter1  & 0.88 & 0.99 & 1.00 & 0.82 & 0.97 & 1.00&0.83 &0.78 & 0.57\\
Twitter2  & 0.22 & 0.99 & 1.00  & 0.05 & 0.88 & 1.00 & 0.56&0.70 & 0.57\\
Twitter3 & 0.74 & 0.98 & 1.00 & 0.04 & 0.65 & 0.99 & 0.05&0.74 & 0.33\\
Slashdot0811  & 0.28 & 0.49 & 0.94& 0.08 & 0.13 & 0.37 &0.09 &0.10 & 0.04\\
\hline
Collab-AstroPhysics  & 0.43 & 0.53 & 0.77 & 0.27 & 0.49 & 0.65& 0.34  &0.83 & 0.71\\
Collab-CondMat  & 0.28 & 0.40 & 0.50  & 0.17 & 0.30 & 0.39& 0.30 &0.71 & 0.52\\
Collab-HighEnergy  & 0.16 & 0.28 & 0.43 & 0.10 & 0.19 & 0.29& 0.19 &0.42 & 0.13\\
Cite-HighEnergy  & 0.13 & 0.35 & 0.55 & 0.15 & 0.31 & 0.47 & 0.30&0.75 & 0.62\\
\hline
Amazon0302& 0.01 & 0.02  & 0.04  & 0.00 & 0.00 & 0.00 & 0.00 & 0.11 & 0.00\\
Epinions  & 0.46 & 0.88 & 0.81  & 0.24 & 0.51 & 0.63 &0.30 &0.18 & 0.12\\
Web-NotreDame  & 0.01 & 0.03 & 0.11 & 0.00 & 0.05 & 0.18 &0.04 &0.07 & 0.03\\
Web-Stanford  & 0.03 & 0.04 & 0.39& 0.00 & 0.09 & 0.46 & 0.02&0.49 & 0.40\\
Wiki-Vote  & 0.48 & 0.96 & 1.00& 0.51& 0.88 &0.99 & 0.51&0.43 & 0.35 \\
\hline
Protein-Collins & 0.00 & 0.16 & 0.93 & 0.00 & 0.79 & 0.95&0.01 &0.59 & 0.36  \\
\hline
\end{tabular}
\vspace{-5px}
\end{center}
\end{table*}

\section{Fundamental Partitionability of Networks}
\label{fundamentalSection}
Some datasets have a higher proportion of cliques split by partitions than others.  This is largely uncorrelated with the mere number of cliques in the dataset, or the number of cliques per node, or per edge, or a number of other simple graph measures, such as clustering co-efficient.
After investigating several popular CFAs, we now consider whether any partition exists which would not split cliques. Perhaps there were potential partitions that would confine cliques to the cores of the communities found, but these methods were not finding them?
To answer this, we consider, for each network, the subgraph induced by nodes that share cliques; i.e. the network formed by discarding all edges from the network that are not part of cliques.
The connected components in this subgraph are the sets of nodes that cannot be placed into separate partitions without splitting any cliques.
We calculate the size of the largest connected component of each network, and present this as the proportion of nodes in the network, in Table \ref{damagedoneGraclus}.

We show results for the subgraph induced by nodes that share cliques of size 4 or greater, and of size 5 or greater, under the headings `4-Clique' and `5-Clique'.
On some networks, such as Facebook, Twitter, or collaboration networks, any partitioning scheme that is constrained to not split cliques of size five or greater has to leave the majority of nodes in a single partition.

This is an important structural property of these datasets, and an important result for certain diffusion models of complex contagion \cite{centola2007complex} which can only spread over structurally strong ties, as it shows these graphs are connected when using solely strong ties -- it is possible to walk the graph communities without using weak ties.
Further, on some of the larger datasets such as the Slashdot dataset, with 77,360 nodes, we find that over 30 per cent of those nodes (23,980 nodes) are in a connected component of the subgraph containing only edges that are in \textit{triangles}; further evidence against the strict idea that strong ties do not cross community boundaries, and that communities are well separated.

\subsection{Partitions that Directly Minimise Clique Splits}
Having established the limits of partitions that break no \textit{single} clique, we consider partitioning to directly optimise the number of cliques preserved, while producing balanced partitions.
Partitioning a network while splitting as few cliques as possible is a hypergraph partitioning problem, where nodes in a clique together are connected by a hyperedge.
This simple observation enables us to use a balanced mincut hypergraph partitioning algorithm, such as implemented by \textit{hMETIS} \cite{karypis1999multilevel} to partition the graph, while directly minimising clique splitting.
\textit{hMETIS} requires an important parameter to determine partition balance.
Too high a value results in trivial partitions, with the vast majority of nodes in a single partition; too low might force \textit{hMETIS} to make more aggressive hyperedge cuts than is reasonable.
We initially set this \textit{ufactor} at 50 (meaning the largest partition may have 50\% larger weight than the average), to allow some unbalance.
We examine cuts into 4, 16, and 64 partitions -- generally fewer partitions than the modularity maximisation approach finds on these graphs.
We also present results for 16 partitions with \textit{ufactor} 500, allowing very large variation in partition size.

The results are shown in Table \ref{damagedoneGraclus}.
Partitions directly minimising clique split indeed result in reduced proportions of the cliques split, compared to the balanced mincut of \textit{Graclus}.
As the number of partitions, and balance between partitions, constrain \textit{hMETIS} more than the modularity maximisation method, the results are not directly comparable.
However, as this method is directly minimising clique cut, it should approach a lower bound attainable by any partitioning CFA, for the given number of partitions -- and, with generous balance parameters, indeed does better than modularity maximisation.

Even so, partitioning the network using this method, on a range of datasets -- notably the collaboration networks, the Wiki voting data, the telecoms data and especially the Facebook social networks -- still results in substantial proportions of cliques being split, demonstrating the fundamental global unpartitionability of some networks.

\subsection{Detailed Analysis of Sample Networks}
\label{detailsection}
We now present some detailed statistics from three arbitrarily chosen sample networks: the Princeton Facebook network, which we will look at in detail as a case study, and one of each of the mobile and twitter sample networks.
This Princeton Facebook network with over 6,500 nodes is large enough to allow us meaningfully investigate medium and large scale community structure.
Facebook network data is also relatively dense, in that it captures many long term social relationships for each user; this is in contrast to more fleeting, or partial, network information we might obtain by extracting a network from a short term snapshot of a communications network.

\begin{figure}[!htb]
\begin{center}
\subfigure[]{
    \includegraphics[width=2.2in]{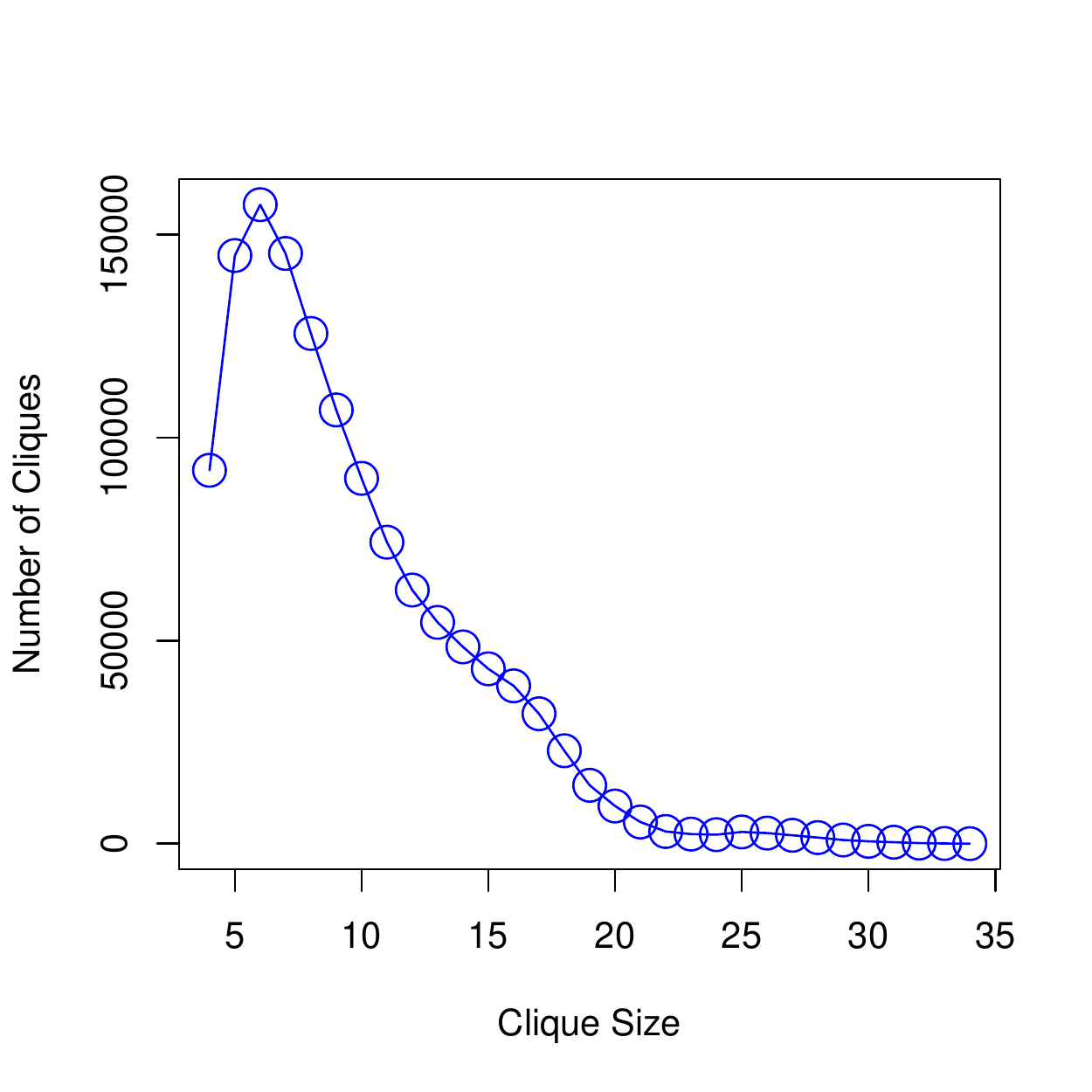}
    \label{CliqueFreqGraphsPrinceton}
}
\subfigure[]{
    \includegraphics[width=2.2in]{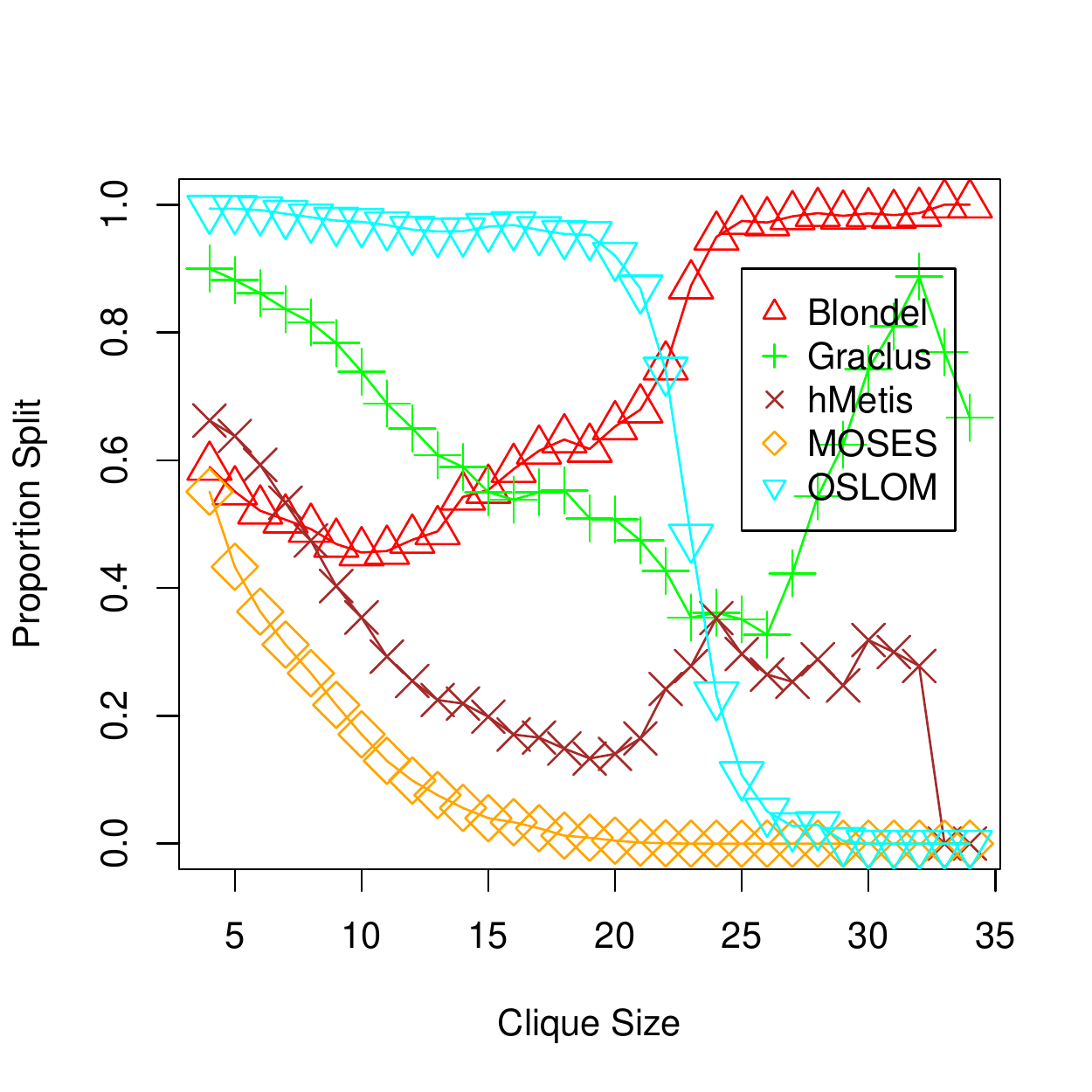}
    \label{CliqueSplitProportionsGraphsPrinceton}
}
\subfigure[]{
    \includegraphics[width=2.2in]{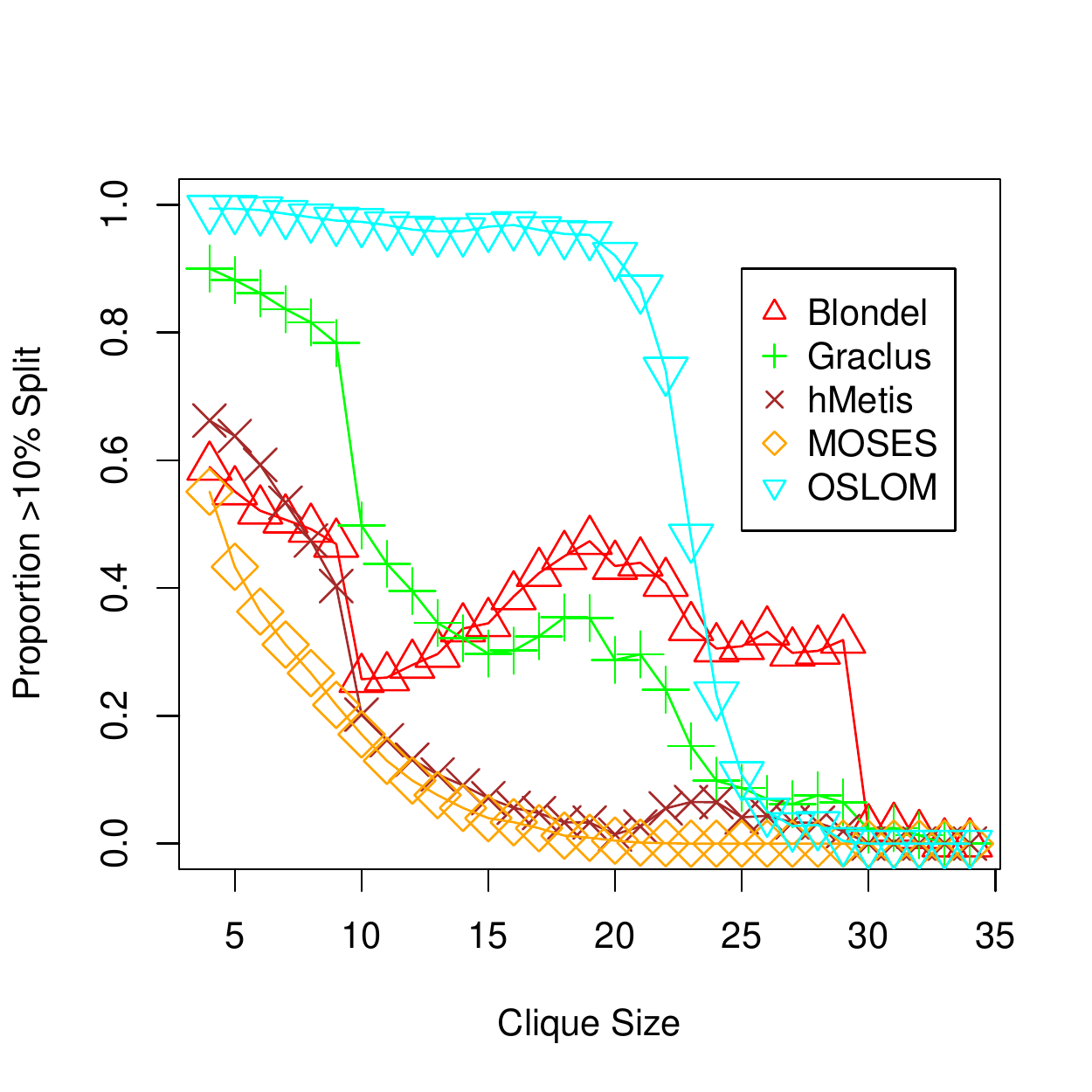}
    \label{CliqueSplitProportionsGraphsPrinceton10pc}
}
\subfigure[]{
    \includegraphics[width=2.2in]{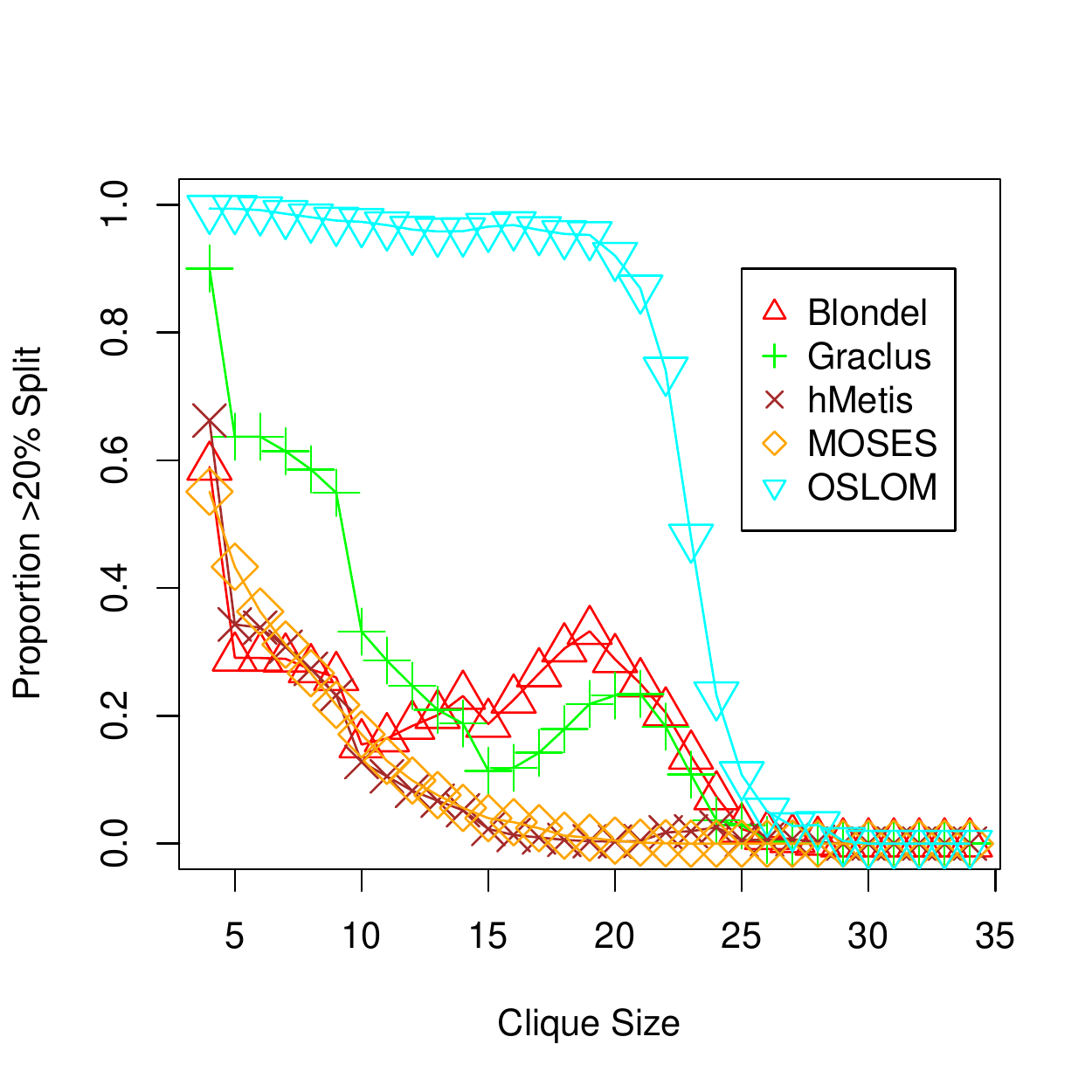}
    \label{CliqueSplitProportionsGraphsPrinceton20pc}
}
\caption{Proportion cliques split at each size, for Princeton Facebook Network (b) Considering a clique split if a single node is partitioned from it (c) Considering a clique split if $>$10\% of its nodes are partitioned from it (i.e.\ $\alpha$ =.9) (d) Considering a clique split if $>$20\% of its nodes are partitioned from it (i.e.\ $\alpha$ =.8). }
\label{detailedViewPrinceton}
\end{center}
\vspace{-4px}
\end{figure}

\begin{figure}[!htb]
\begin{center}
\subfigure[]{
    \includegraphics[width=2.2in]{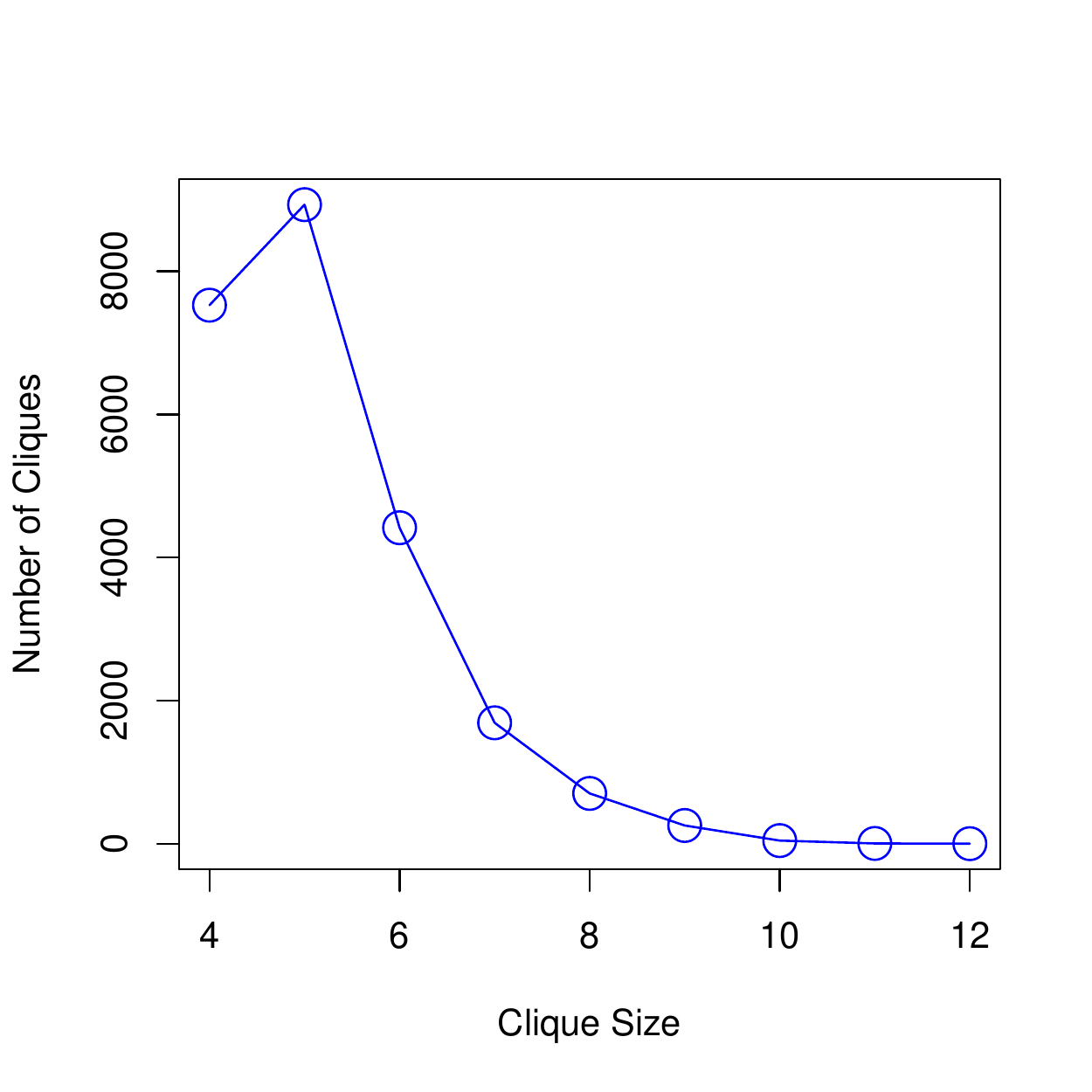}
    \label{CliqueFreqGraphsTwitter}
}
\subfigure[]{
    \includegraphics[width=2.2in]{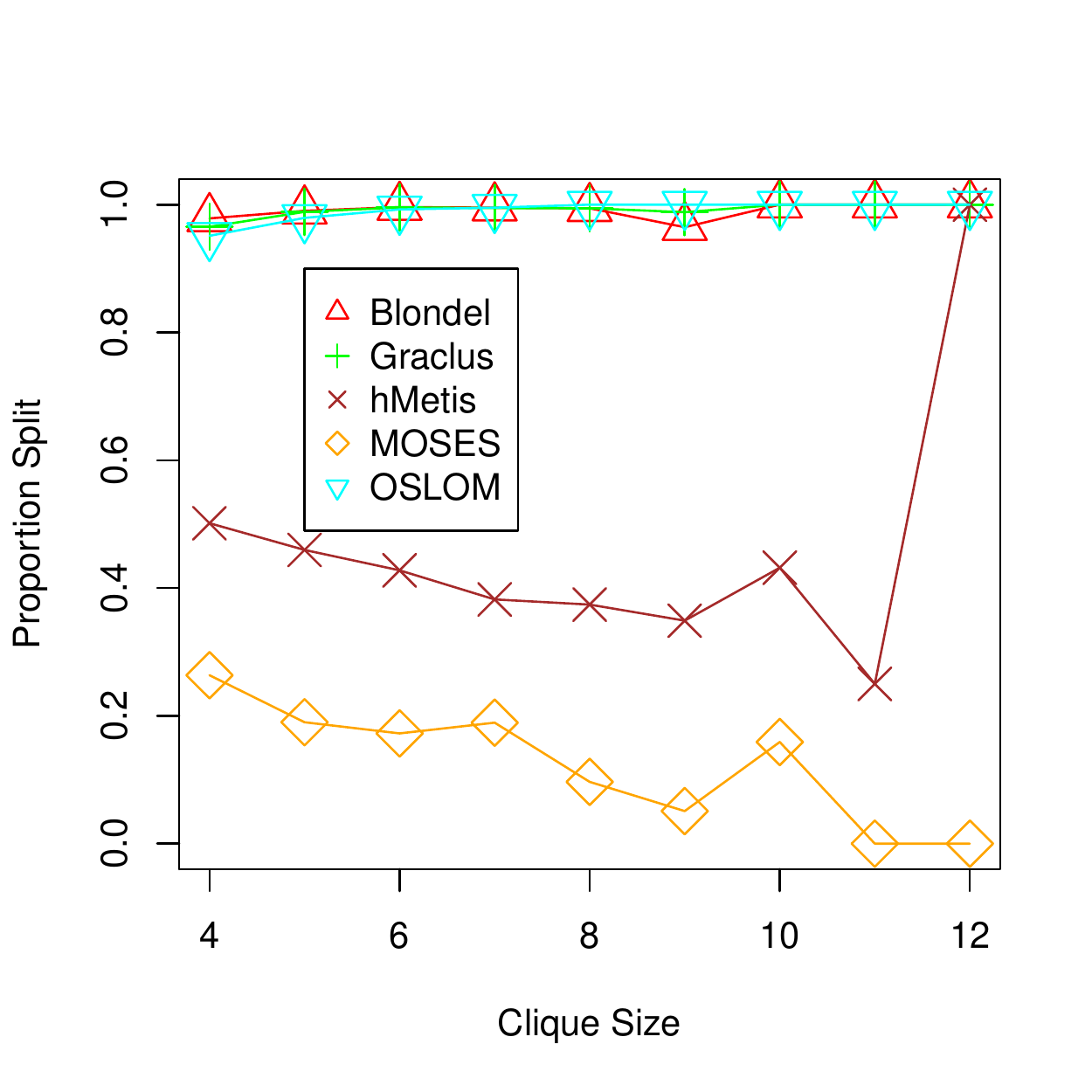}
    \label{CliqueSplitProportionsGraphsTwitter}
}
\subfigure[]{
    \includegraphics[width=2.2in]{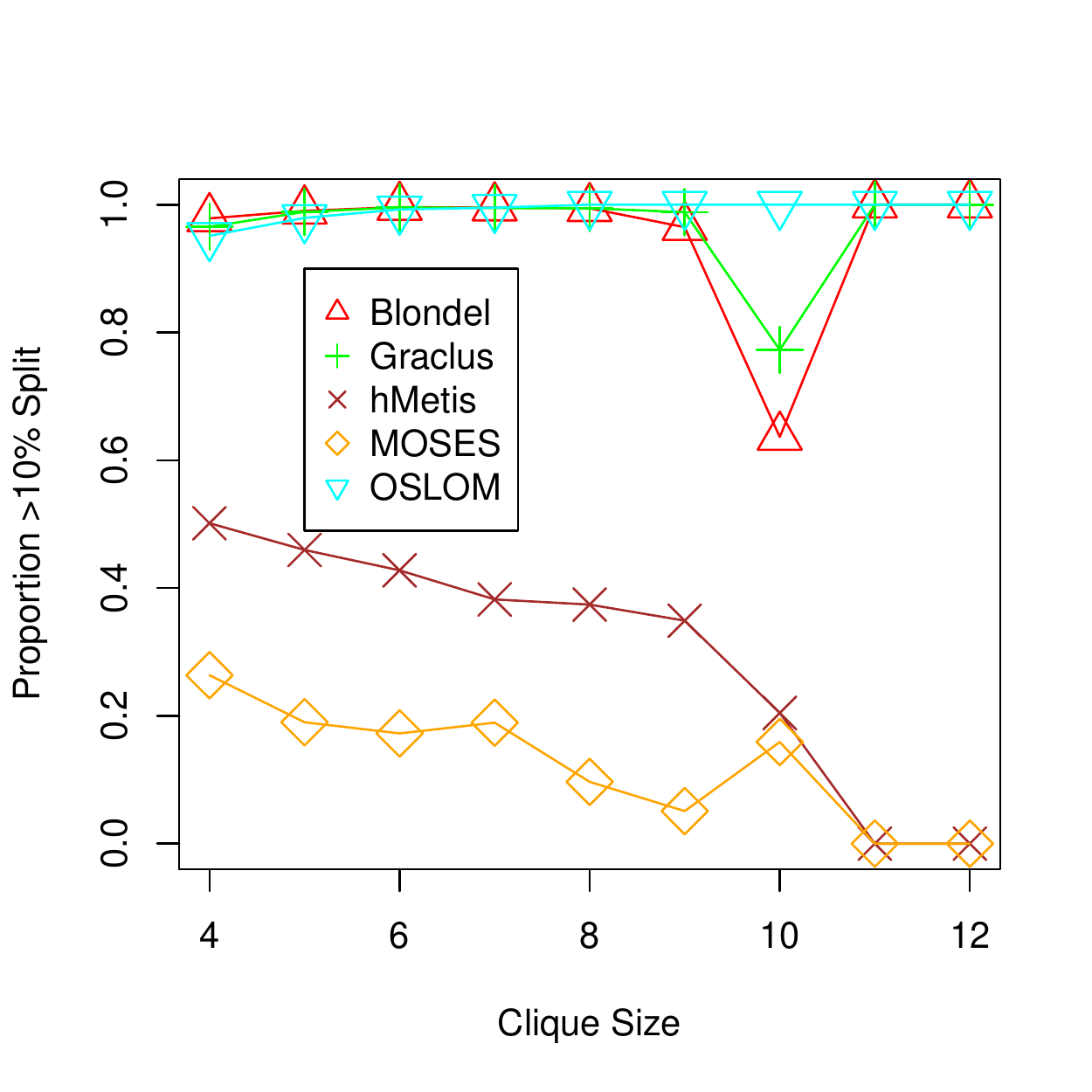}
    \label{CliqueSplitProportionsGraphsTwitter10pc}
}
\subfigure[]{
    \includegraphics[width=2.2in]{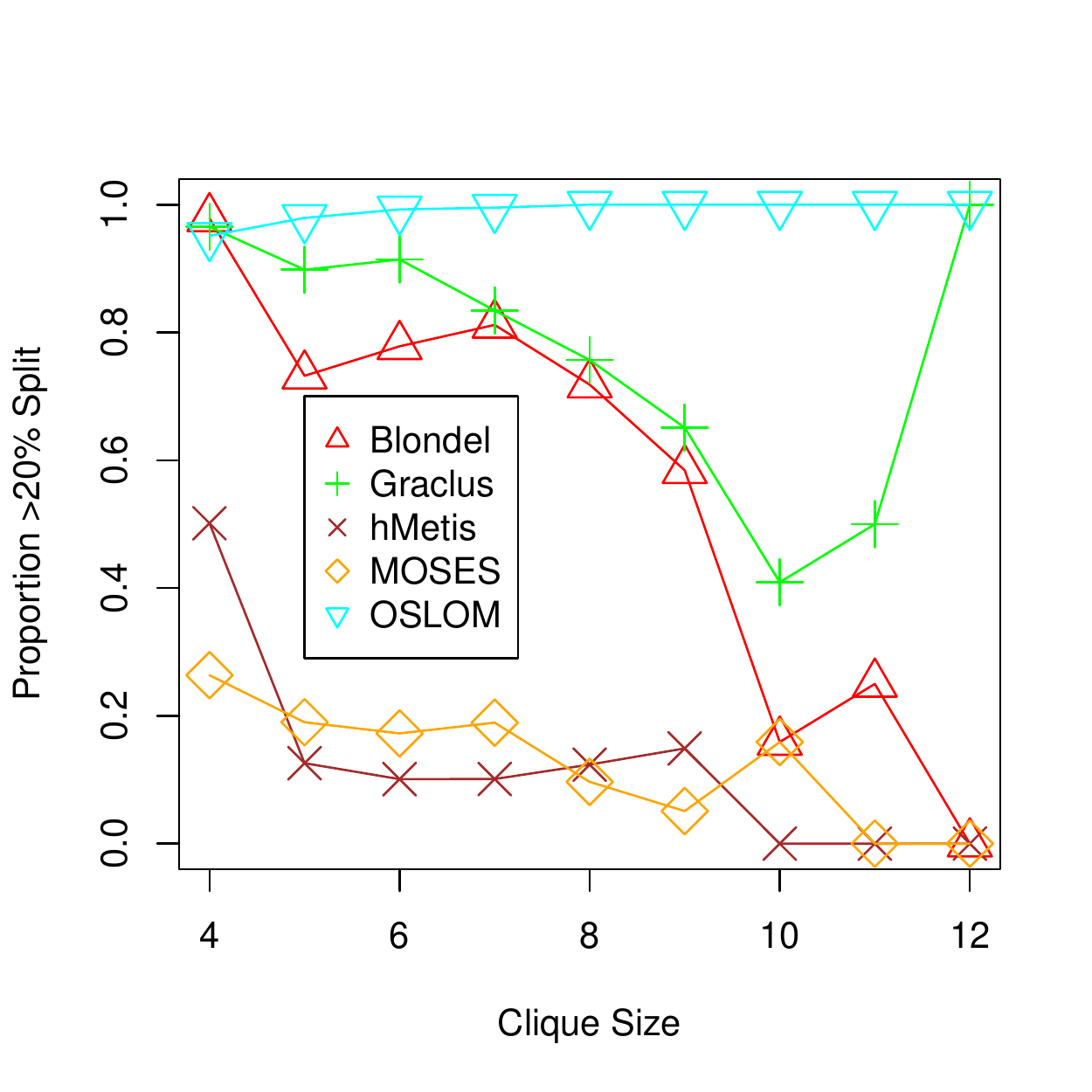}
    \label{CliqueSplitProportionsGraphsTwitter20pc}
}
\caption{Proportion cliques split at each size, for one of the Twitter Networks (b) Considering a clique split if a single node is partitioned from it (c) Considering a clique split if $>$10\% of its nodes are partitioned from it (i.e.\ $\alpha$ =.9) (d) Considering a clique split if $>$20\% of its nodes are partitioned from it (i.e.\ $\alpha$ =.8). }
\label{detailedViewTwitter}
\end{center}
\vspace{-4px}
\end{figure}

\begin{figure}[!htb]
\begin{center}
\subfigure[]{
    \includegraphics[width=2.2in]{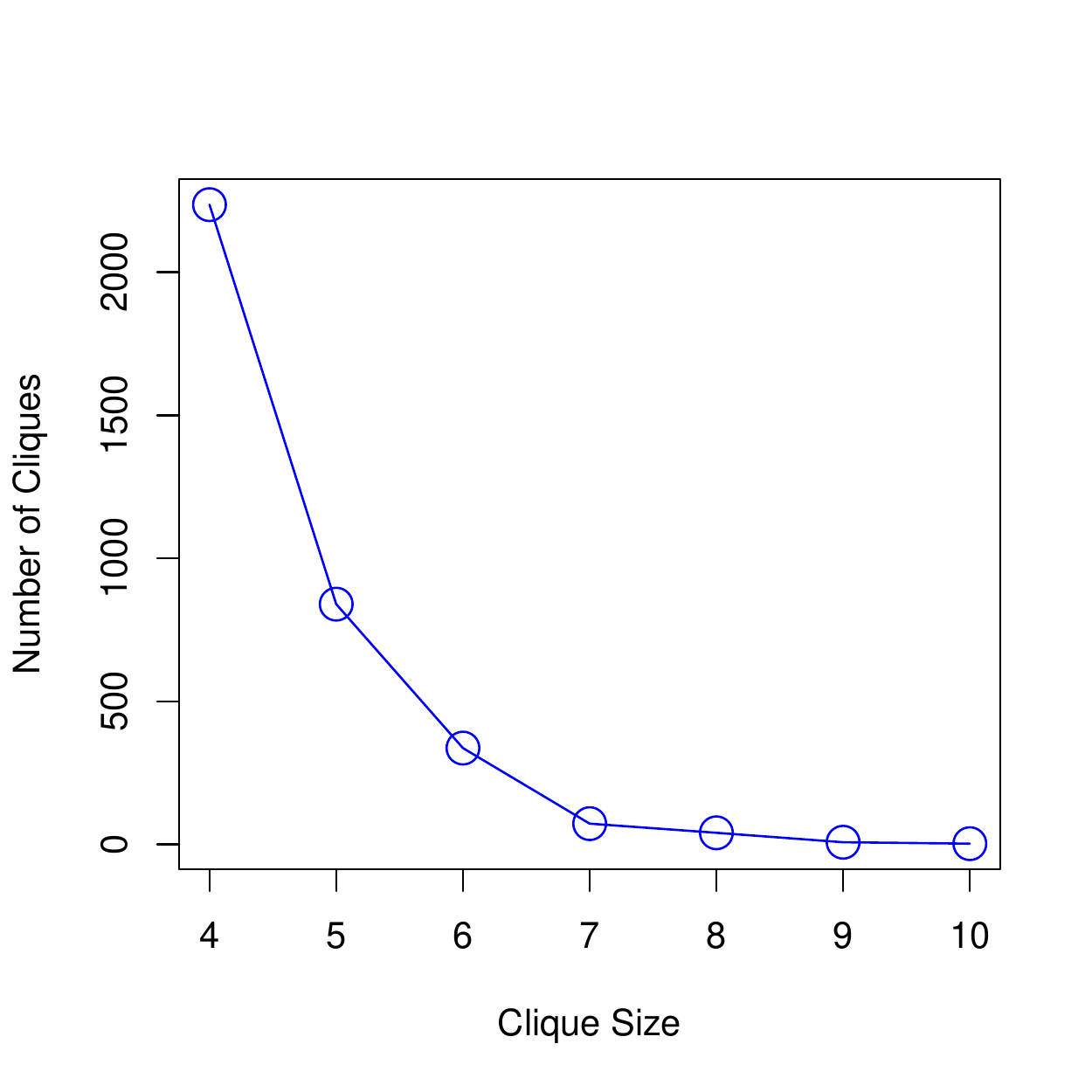}
    \label{CliqueFreqGraphsMobile}
}
\subfigure[]{
    \includegraphics[width=2.2in]{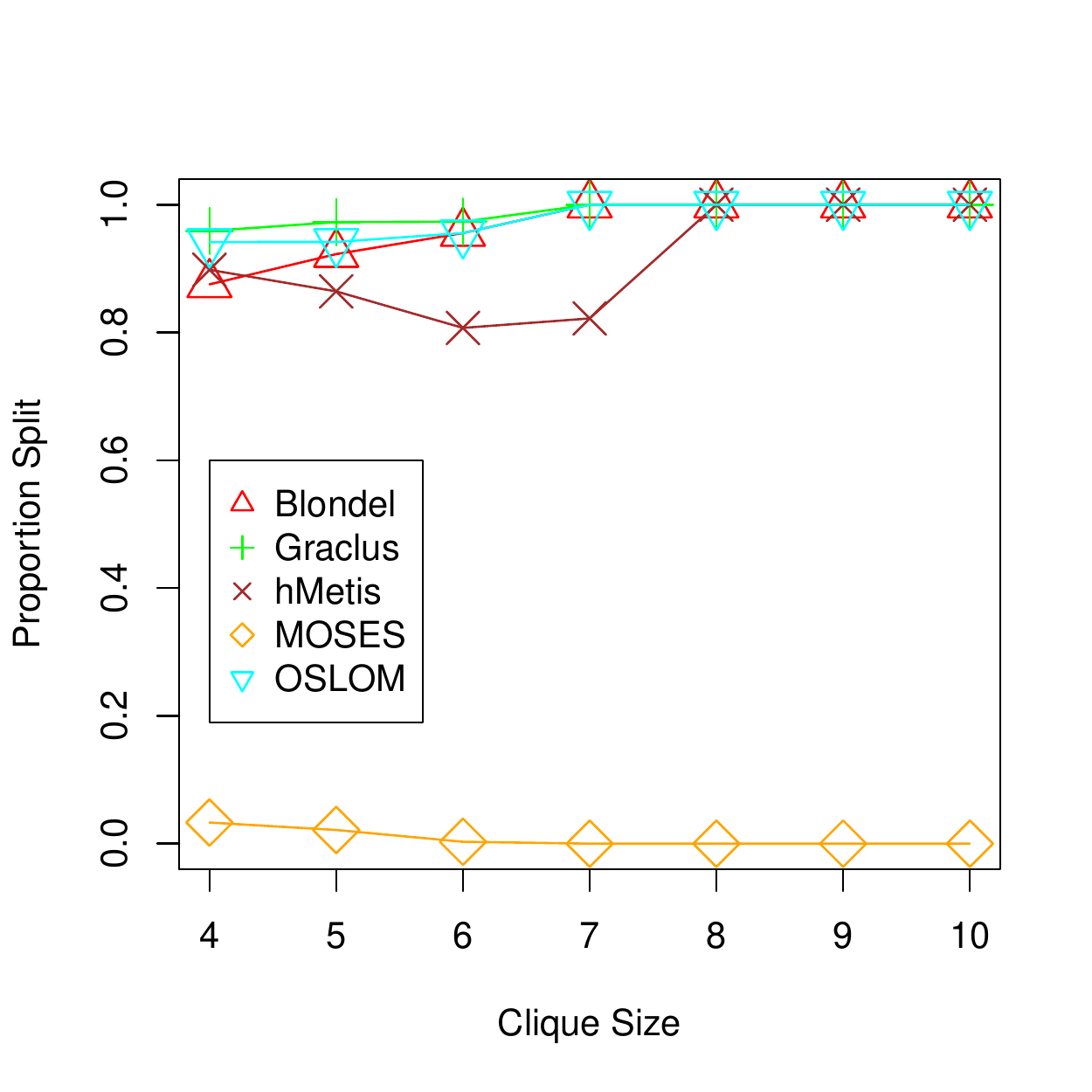}
    \label{CliqueSplitProportionsGraphsMobile}
}
\subfigure[]{
    \includegraphics[width=2.2in]{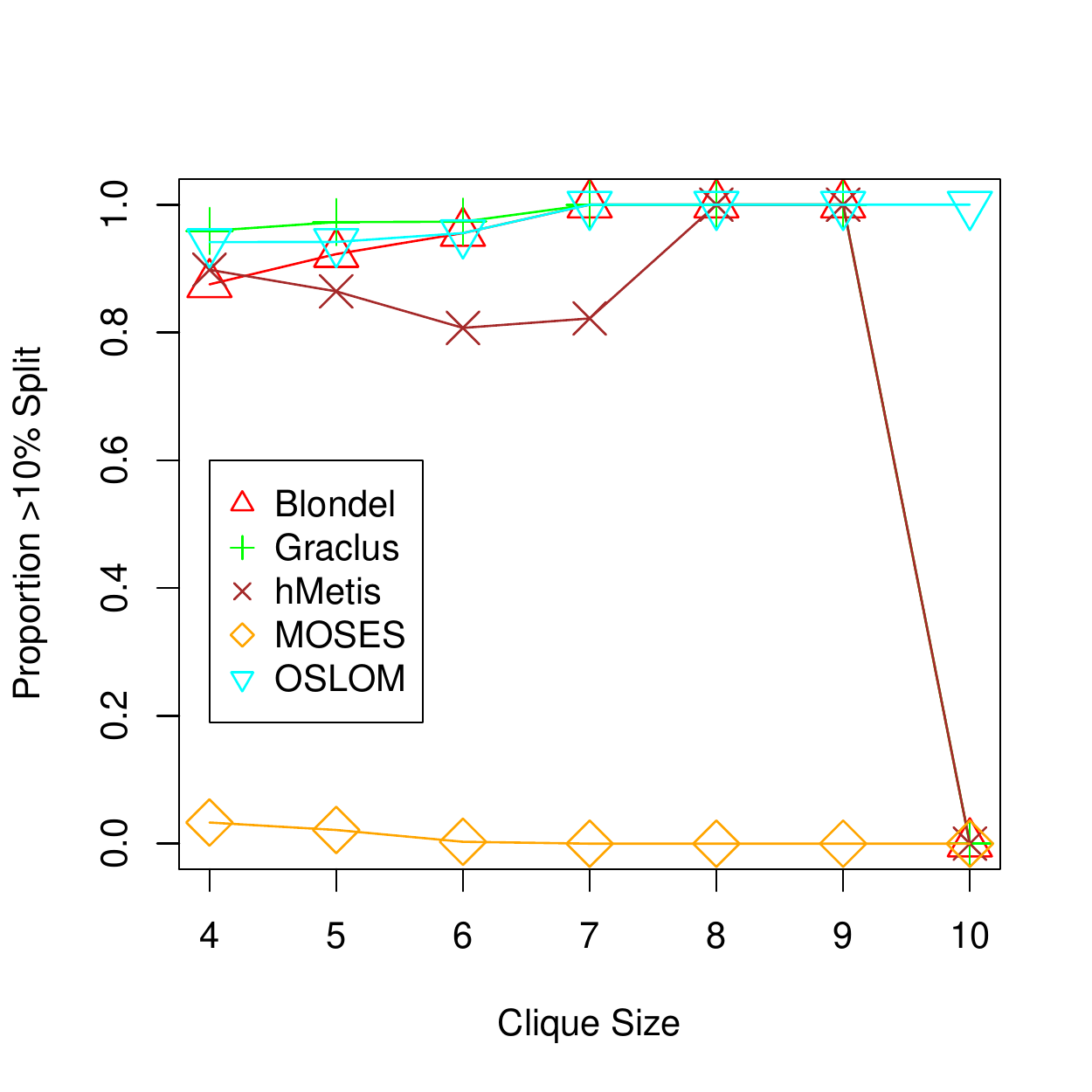}
    \label{CliqueSplitProportionsGraphsMobile10pc}
}
\subfigure[]{
    \includegraphics[width=2.2in]{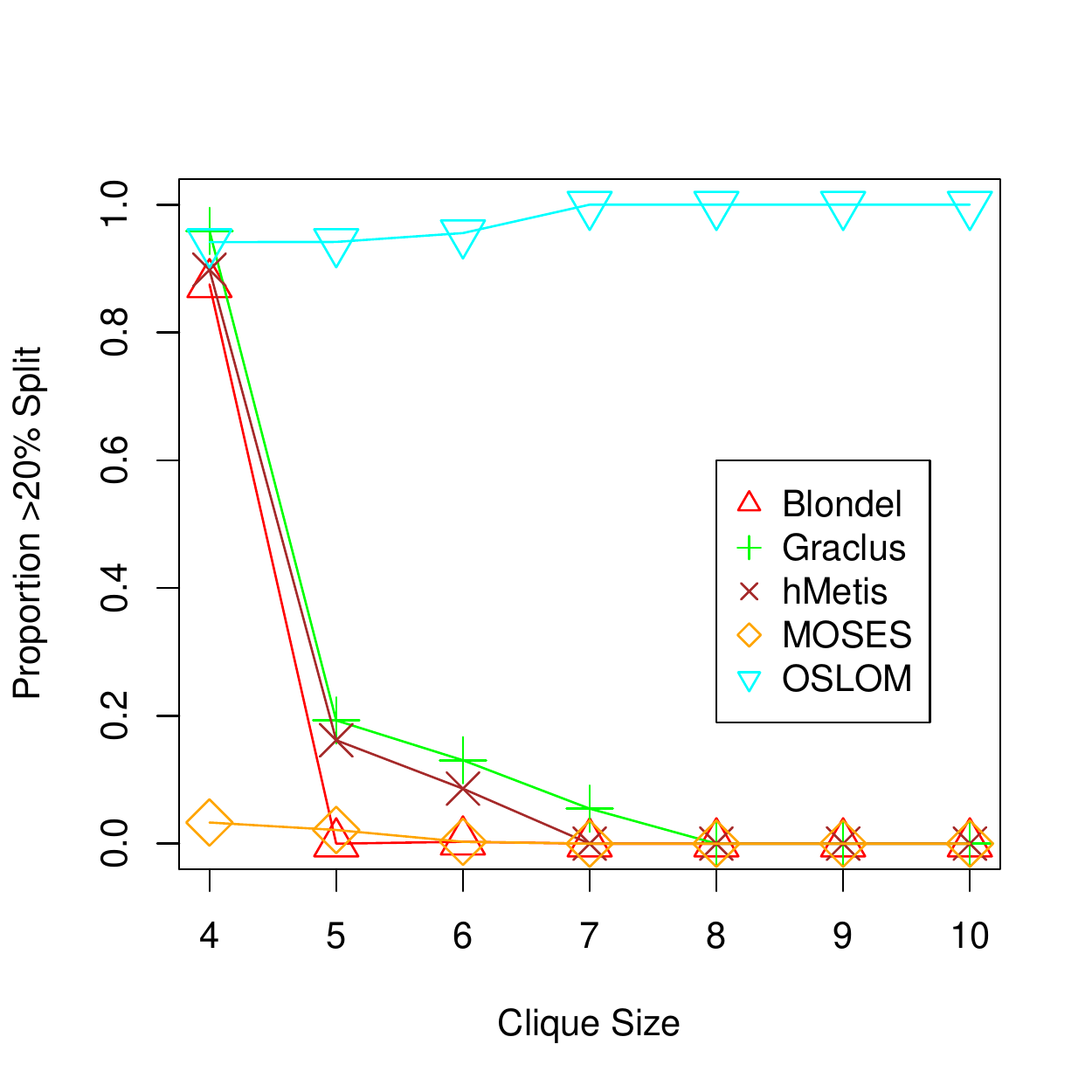}
    \label{CliqueSplitProportionsGraphsMobile20pc}
}
\caption{Proportion cliques split at each size, for one of the Mobile Networks (b) Considering a clique split if a single node is partitioned from it (c) Considering a clique split if $>$10\% of its nodes are partitioned from it (i.e.\ $\alpha$ =.9) (d) Considering a clique split if $>$20\% of its nodes are partitioned from it (i.e.\ $\alpha$ =.8). }
\label{detailedViewMobile}
\end{center}
\vspace{-14px}
\end{figure}

In Figure \ref{detailedViewPrinceton} we show the number of cliques of each size in the network.
We also show, for each clique size $n$ observed in the network, the number of split cliques of that size.
We plot this profile of cliques split, at each size, for each partitioning method investigated (as well as for non-partitioning Overlapping Community Finding Algorithms, which will be discussed in Section \ref{overlappingCommunitySection}).
We also show the proportion of cliques of size $n$ split, for each value of $n$.
We present results for three definitions of `split' -- where we consider cliques split if (b) any of their nodes have been partitioned from them, (c) greater than 10\% of their nodes have been partitioned from them, and (d) greater than 20\% of their nodes have been partitioned from them.

The Louvain method finds 21 partitions on this network; we use \textit{Graclus} and \textit{hMETIS} to produce the same number of partitions as the Louvain method.
While the absolute number of cliques split tends to decrease as the metric becomes increasingly conservative, we note that in all cases, non-trivial numbers of cliques are split.
As we would expect -- as all partitioning algorithms try, in some sense, to avoid cutting edges -- often the larger a clique is in size, the smaller the probability of the partitioning algorithms splitting it; however, cliques of all sizes are still split by these methods, on some networks; it is not the case that only the smallest cliques are split.

These figures emphasize the robustness of our findings -- cliques of all sizes are split by partitioning -- and illustrate an interesting way of characterising the effects of partitioning a network.
Figures \ref{detailedViewTwitter} and \ref{detailedViewMobile} show similar results for one of each of the Twitter and Mobile networks.

\subsection{`Distinct' Cliques}

A large clique, with some small portion of random edges deleted, will turn into many very similar smaller cliques.
In quantifying the `proportion' of cliques split, we might be concerned that mis-assignment of a small set of nodes, if they are contained within a large number of very similar overlapping cliques, might skew the proportions.
As an additional check on the robustness of these results, we present a set of results in Table \ref{damagedoneJaccard} which correct for this effect by running our analysis not on the full set of maximal cliques, but instead on a set of maximal cliques, after a pre-processing phase which removes any clique that has a high Jaccard similarity ($>$0.8) to any other larger clique.
This analysis is computationally expensive to compute on the larger networks; however, on the networks we are able to perform it on, we find that our results still hold: substantial proportions of cliques are split, even if the only cliques we are looking at are cliques that are somewhat distinct from each other.
\begin{table}[ht]
\caption{Proportion of cliques that are `distinct', beyond a given Jaccard similarity, that are over 10\% split (i.e.\ $\alpha$=.9) by \textit{Graclus} \cite{dhillon2007weighted}, and \textit{hMETIS} \cite{karypis1999multilevel}. Values shown for \textit{Graclus} and \textit{hMETIS} for 16 partitions. \textit{hMETIS} \textit{ufactor} is 500.}
\label{damagedoneJaccard}
\begin{center}
\begin{tabular}{r|rrr}
\hline
Network & Louvain & Graclus & hMETIS\\
\hline
\hline
Email-Enron  & 0.62 & 0.76 & 0.39 \\
Mobile1  & 0.97 & 0.88 &0.80\\ 
Facebook-caltech  & 0.78 & 0.90 &0.66 \\
Twitter1  & 0.99 & 0.99 & 0.83 \\
Collab-HighEnergy  & 0.23 & 0.28 & 0.19 \\
Protein-Collins & 0.34 & 0.33 & 0.02  \\
\hline
\end{tabular}
\end{center}
\vspace{-17px}
\end{table}

\subsection{Random and Synthetic Models of Community}

\begin{figure}[!htb]
\centering
\vspace{-28px}
\includegraphics[width=2.8in]{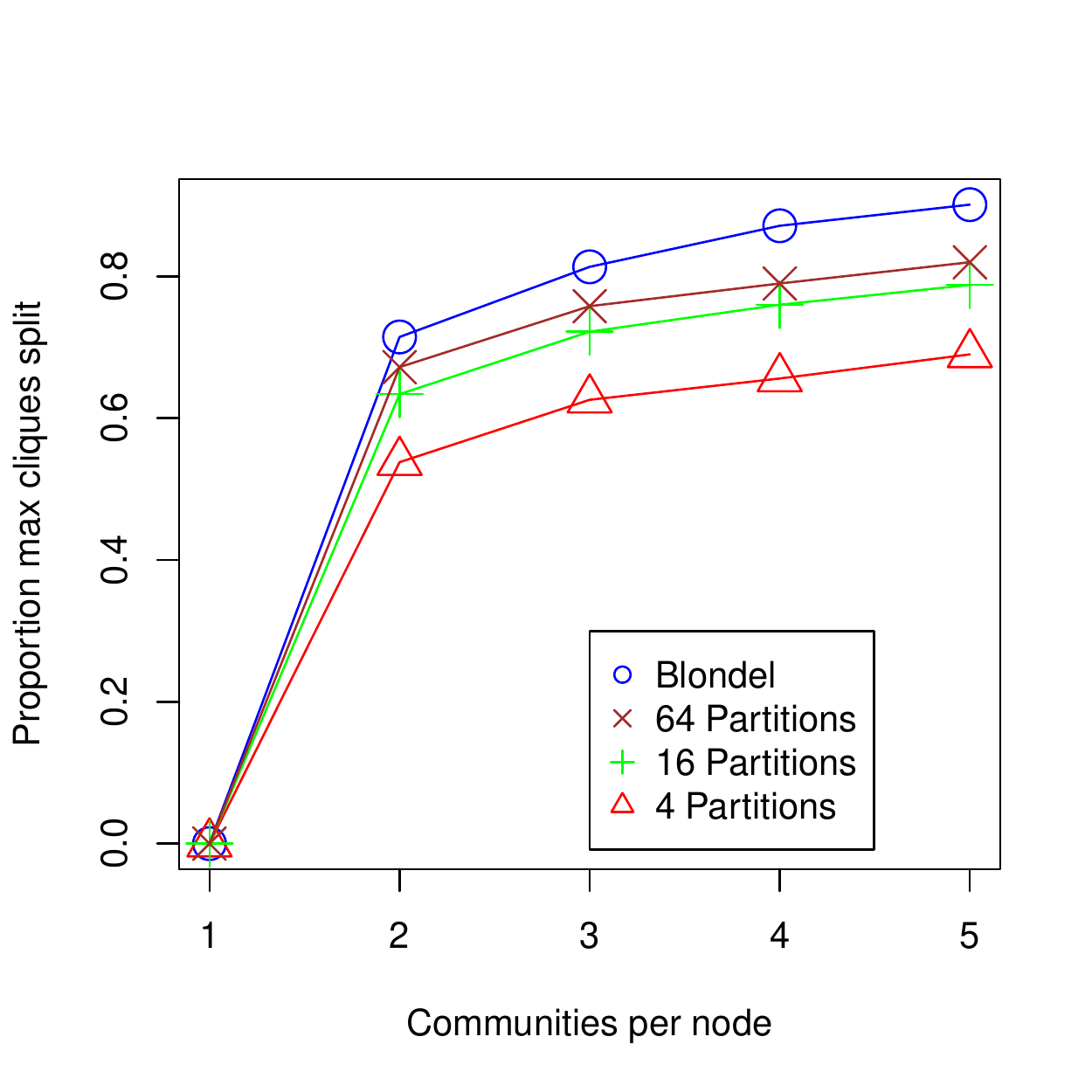}
\caption{Number of communities-per-node, as specified in benchmark parameters, vs proportion of maximal cliques $>$10\% split (i.e.\ $\alpha$ =.9), by the Louvain and hypergraph partitioning methods, on LFR benchmark data. Each data point is the mean of 5 LFR instances; deviation is negligible.}
\label{LFRplot}
\end{figure}

Broad categories of random community assignment model will produce networks where partitioning will fail to recover full communities.
One source of synthetic benchmark community data is the `LFR' benchmark \cite{lancichinetti2009benchmarks}, in which `communities' -- defined as sets of nodes with a high probability of edges between them -- are embedded into a generated network.
We ran our experiments on LFR graphs to test our method on synthetic data.
We generated realisations of a 10,000 node network, altering the number of communities each node was assigned to -- from one to five, also increasing the corresponding number of edges, using the same parameters as with benchmarks detailed in previous work \cite{lee2010detecting}.
The results detailing the proportion of cliques split are shown in Figure \ref{LFRplot}.

From these results, all methods partition the single-community-per-node networks without splitting cliques, but split significant numbers of cliques on networks with two or more communities-per-node. %
Even though the synthetic network model isn't directly embedding cliques -- just increasing edge density within communities -- partitioning fails to find all structure, by our defined metrics, on synthetic networks where nodes are overlapping.
Further, large components exist in the graph of edges in cliques in these generated networks.
Not only are the individual nodes and communities overlapping as designed by the model; it is a \textit{global} property of the network as a whole that \textit{no} non-trivial partition exists which does not split cliques.

\section{Overlapping Community Finding Algorithms}
\label{overlappingCommunitySection}
A variety of algorithms exist which find overlapping communities within networks; Fortunato \cite{fortunato2009community} mentions several of these, but this is an active area of research, with new algorithms frequently being developed \cite{xie2011overlapping}.
Like with partitioning CFAs, many of these algorithms find subtly different structures, as authors work from slightly differing assumptions as to what constitutes a `good' community.
As such it is difficult to interpret what the output of a specific overlapping community finding algorithm tells us about the fundamental structure present in a network; and so we have avoided this in our analysis thus far.
However, an advantage of overlapping CFAs is that they generally find structures that are much less common than maximal cliques are; typically a single overlapping community will contain many maximal cliques, many of which may differ only by a small number of nodes.
As we have seen, a great number of cliques exist in the networks we examine; and while we can investigate aspects of network structure by using these cliques, the fact that so many of them exist brings some disadvantages; specifically, the `clique graph' -- the graph of cliques that overlap each other -- is typically too large to work with.

In previous sections, we have considered the partitionability of networks, concentrating our analysis solely on cliques as the cores of community structure.
The notion of cliques as community cores can be explicitly encoded in a community finding algorithm, both to produce communities that are disjoint, if disjoint cliques are enforced \cite{yan2009detecting} or overlapping, if this criterion is relaxed \cite{palla2005uncovering}, \cite{shen2009detect}, \cite{havemann2010identification}. %
Indeed, this is the approach of a family of overlapping CFAs, which use cliques as `seeds' for communities, including the `Greedy Clique Expansion' (GCE) algorithm \cite{lee2010detecting}, to which the authors of this work have contributed.
The GCE method starts with maximal cliques as seeds and grows these seeds into communities using a local community quality measure.
Thus, it will trivially produce communities in which there exists, for each maximal clique, at least one community that fully contains it.
However, many other overlapping CFAs have kept with an `edge density' notion of community quality, and find communities without any explicit modelling of cliques.
Given the large numbers of cliques present in empirical networks, approaches that do not explicitly model cliques can have computational advantages over those that do.
It is thus interesting to apply our clique based analysis to such algorithms and ask if density-driven community finding algorithms preserve clique cores, when communities are allowed overlap.

In this section, we will make use of overlapping CFAs for two separate purposes.
First, we analyse two overlapping CFAs with the same procedure as the partitioning CFAs, in order to ascertain the extent to which they split cliques in practice.
Second, we examine the \emph{community overlap graphs} created by these CFAs, and use the results of these graphs to examine the effects of partitioning on these networks, given the community structure as found by these particular overlapping CFAs.

\subsection{Algorithms Examined}
\begin{figure}[!htb]
\begin{center}
\subfigure[MOSES Princeton]{
    \includegraphics[width=2.0in]{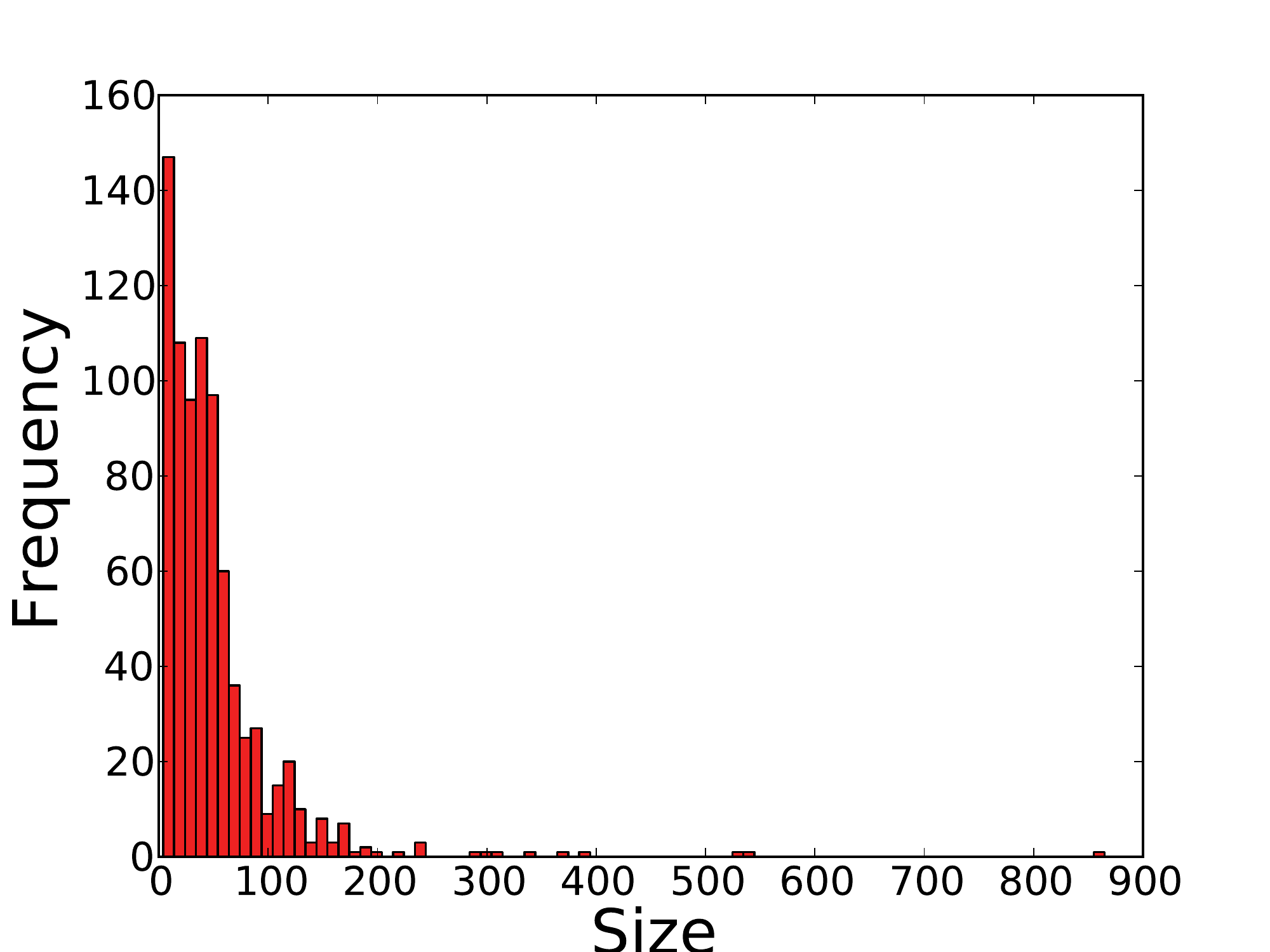}
    \label{CommDistributionMOSESPrinceton}
}
\subfigure[OSLOM Princeton]{
    \includegraphics[width=2.0in]{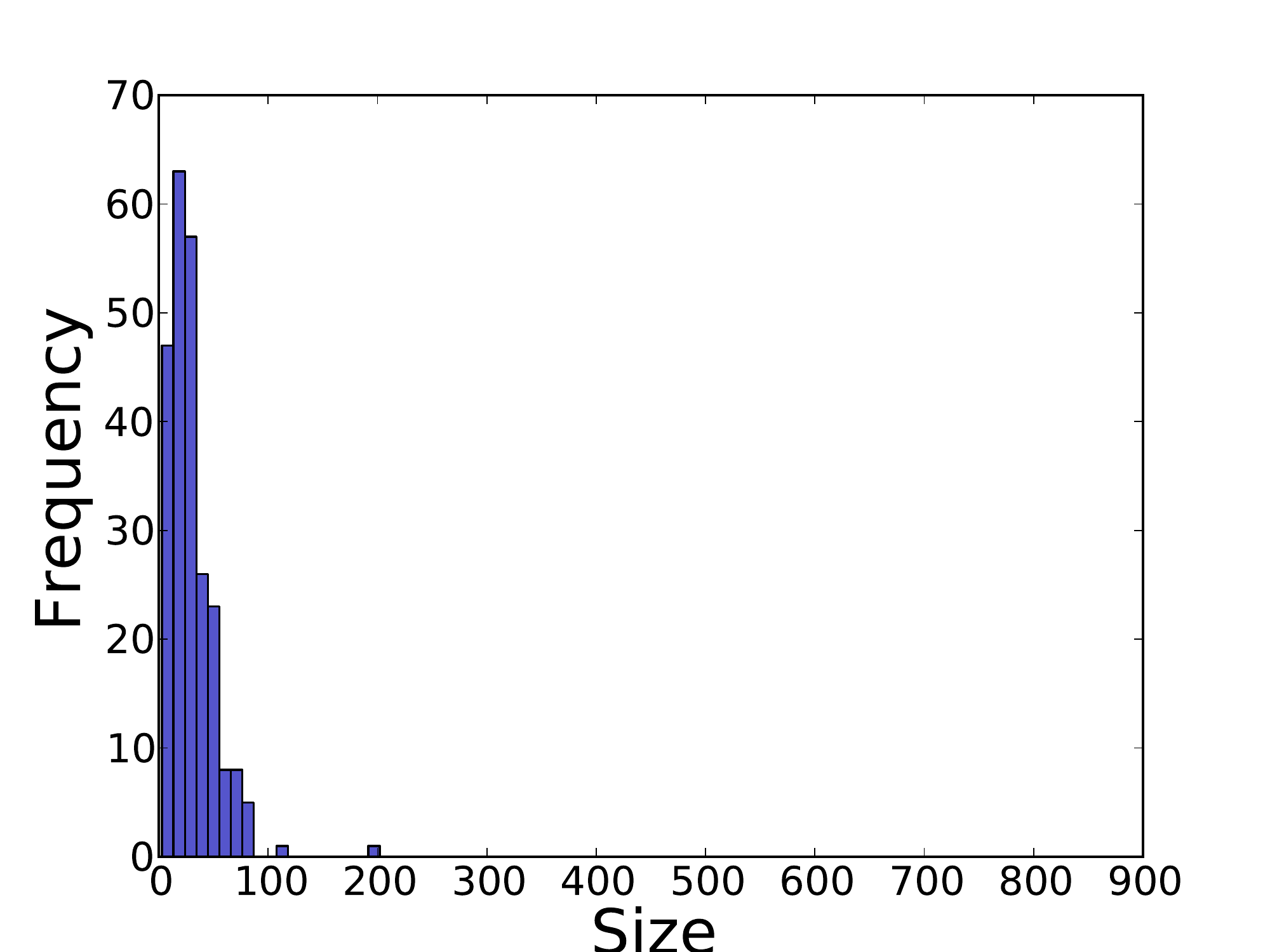}
    \label{CommDistributionOSLOMPrinceton}
}
\subfigure[MOSES Mobile2]{
    \includegraphics[width=2.0in]{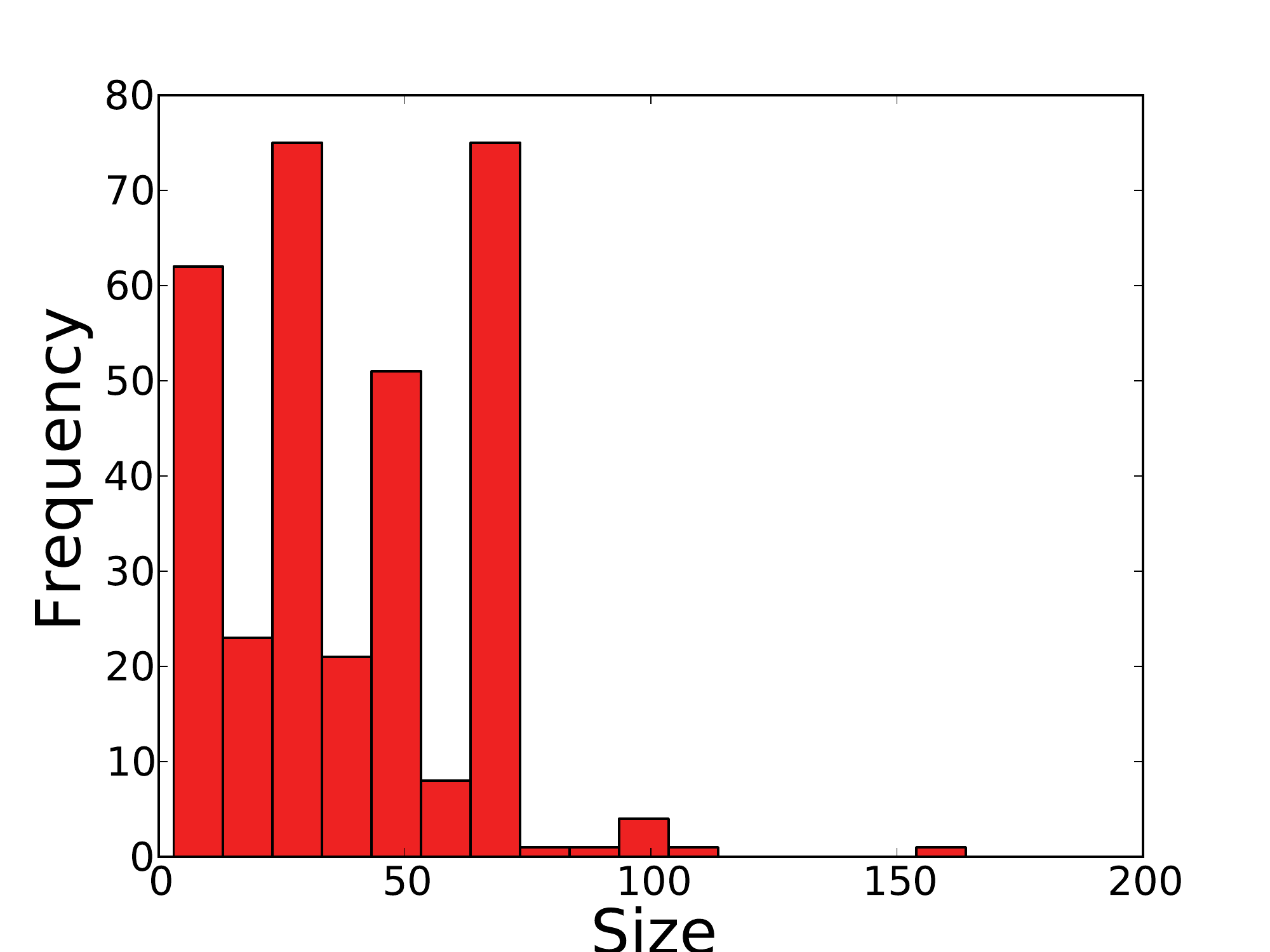}
    \label{CommDistributionMOSESMobile}
}
\subfigure[OSLOM Mobile2]{
    \includegraphics[width=2.0in]{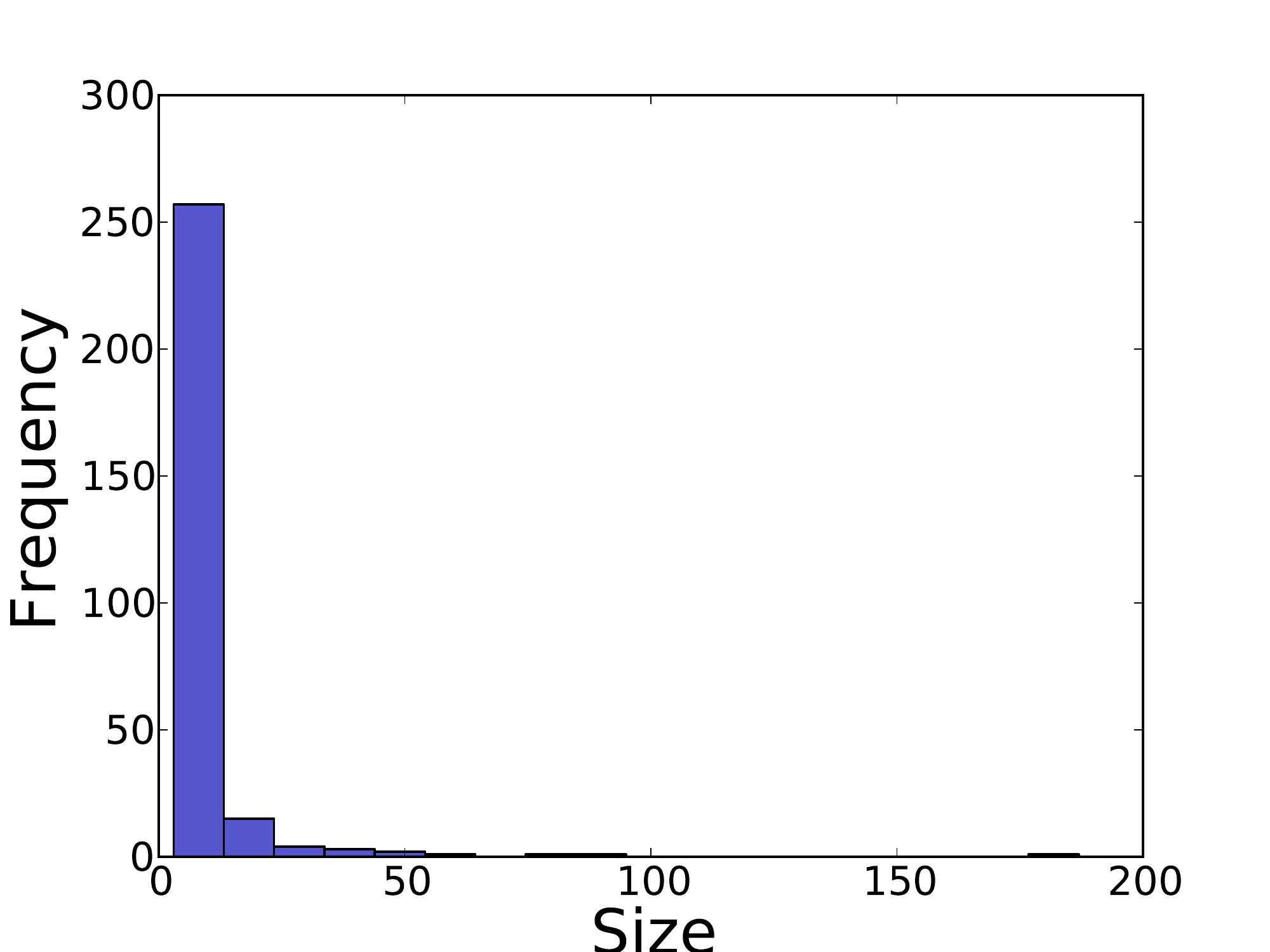}
    \label{CommDistributionOSLOMMobile}
}
\subfigure[MOSES Twitter1]{
    \includegraphics[width=2.0in]{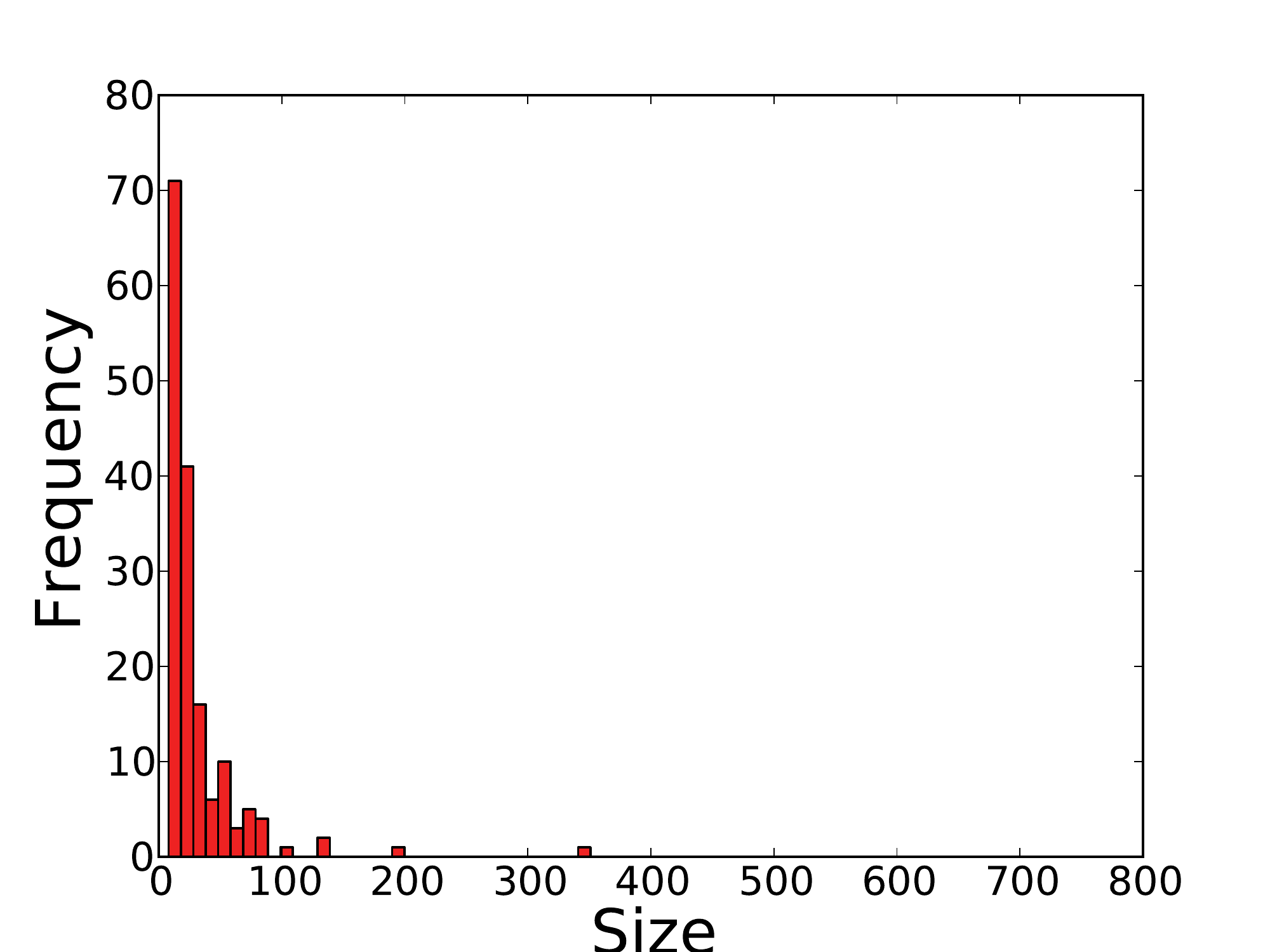}
    \label{CommDistributionMOSESTwitter}
}
\subfigure[OSLOM Twitter1]{
    \includegraphics[width=2.0in]{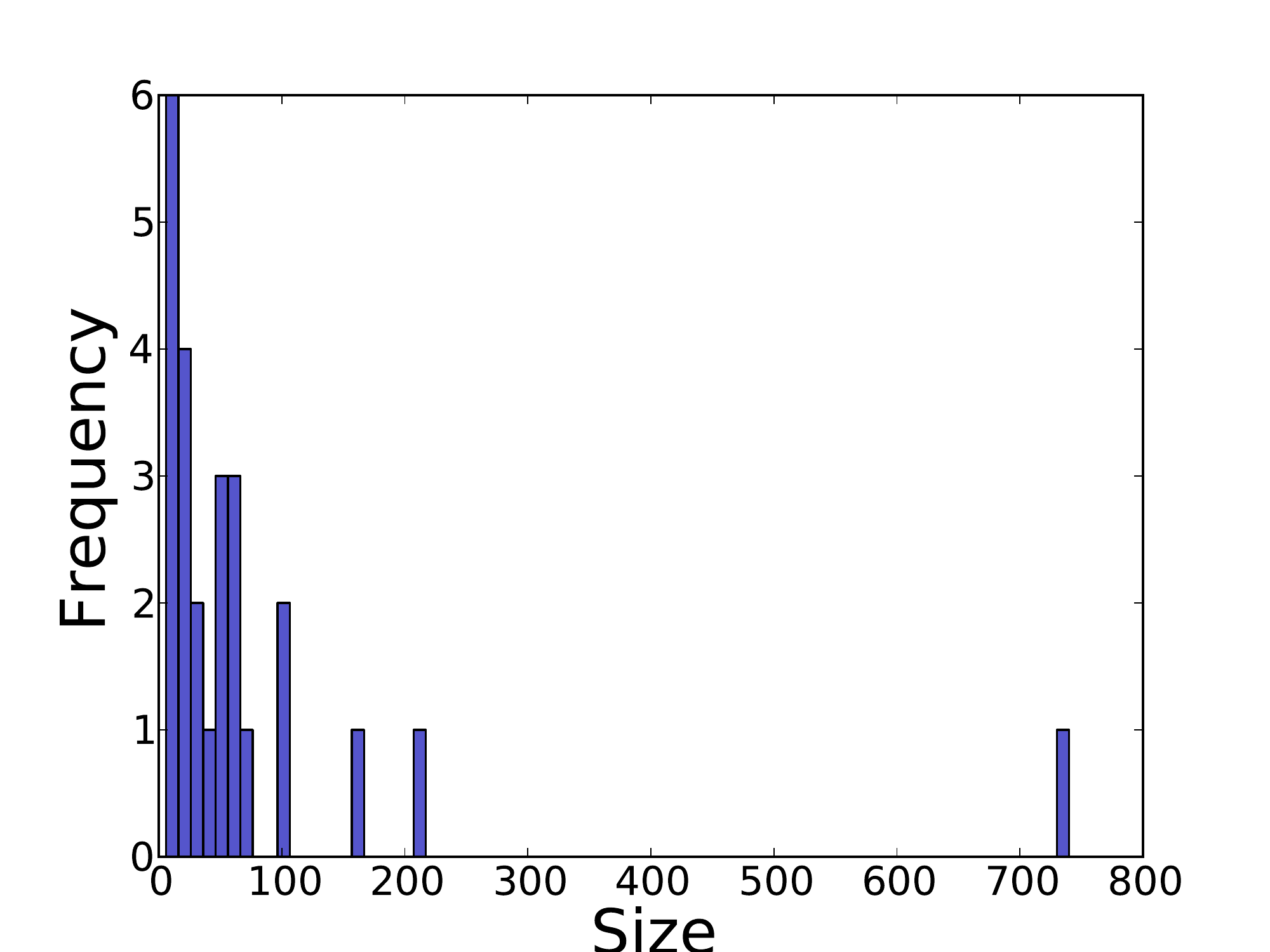}
    \label{CommDistributionOSLOMTwitter}
}
\caption{Size distribution of overlapping communities found by MOSES and OSLOM. We do not show communities consisting of isolated nodes -- OSLOM in particular finds a great many of these.}
\label{commDists}
\end{center}
\vspace{-4px}
\end{figure}

We concentrate our analysis on two recent overlapping CFAs: MOSES (\emph{Model-based Overlapping Seed Expansion}) which we proposed in \cite{mcdaid2010detecting} and OSLOM (\emph{Order Statistics Local Optimization Method}) \cite{lancichinetti2011finding}.
These statistically motivated algorithms do not explicitly use cliques in their operation, and are finding recent application in network analysis, for example the work of Grabowicz et al. \cite{grabowicz2012social}.
While the complexity of these algorithms is dependent on the structure present in the input networks, like the previously discussed methods, both of these algorithms provide implementations which use heuristic techniques to enable them to scale to large networks, this makes them suitable for our analysis.
We first examine the outputs of these algorithms.
Figure \ref{commDists} shows the community size distributions for each of the CFAs we analyze, on the case study networks we presented in detail earlier.
We present distributions, rather than average community sizes, as the distributions of community sizes found vary widely by network, and tend to be heavily skewed.
We do not include `singleton' communities, containing only single nodes; OSLOM in particular reports many of these.

\subsection{Analysis in terms Of Split Cliques}
We analyzed the overlapping communities found by OSLOM and MOSES, in the same manner as the partitioning algorithms -- for each clique, we check to see if there is any community in which it is fully contained; if there is not, we consider the clique to be split.

A thorough comparison of the exact structures found by these algorithms, each motivated by slightly differing models of community structure, is outside the scope of this work.
To be thorough, we would have to either deal in subtle differences in the definition of `community' -- for which many definitions exist -- or analyze the communities found by these methods against some `ground-truth' data particular to a specific application domain.
To restrict our analysis solely to network structure, we do not consider the issue of whether the communities found by MOSES and OSLOM are overall `good' communities; instead we maintain our focus on split cliques.
We present the results of this analysis in Table \ref{damageDoneOverlapping}.
We also show detailed results, for our case-study networks, on a per size-of-clique basis in Figures \ref{detailedViewPrinceton}, \ref{detailedViewTwitter} and \ref{detailedViewMobile}.
These results show that MOSES produces a set of communities such that most larger cliques, in most networks, are contained in at least one community found by MOSES. 
This is an interesting result, considering that MOSES does not explicitly find communities in terms of cliques.
The benchmarking of OSLOM yields a different result, however: for large numbers of cliques, OSLOM does not produce at least one community containing the clique.

It is difficult to explain this. %
Unlike MOSES, OSLOM outputs a hierarchy of levels of community, and we only consider the lowest level of that hierarchy; perhaps, in practice, the lowest levels are very `fine grained' for OSLOM, below the level of an individual clique.
Alternatively, in any community finding algorithm, there must always be a tradeoff between the sensitivity required to find all communities, and specificity to avoid finding `false positive' communities.
We can use cliques as underestimates of community structure, to measure sensitivity -- in that every clique should be contained in a community -- but not to measure specificity, for as discussed earlier, it may be too strict to require that every community contain a clique.
Perhaps OSLOM is simply more specific in its output than MOSES; the two algorithms find structures of different quantity and size, as Figure \ref{commDists} shows.
A detailed investigation of these issues would have to be undertaken in the context of a specific application domain, with ground truth data. But what these results do show is that at least some overlapping CFAs, which contain no explicit representation of cliques, find communities which split much fewer cliques than the partitioning algorithms do.
The results also show that while the use of an overlapping CFA is necessary to avoid splitting cliques, as discussed in Section \ref{fundamentalSection} and concretely illustrated by the results of MOSES, it is not a sufficient condition in practice as the OSLOM results show.
Thus if a specific application domain requires high sensitivity, and a full list of community structure to be found, then not only must an overlapping CFA be used, but the CFA should also be evaluated for the specific application.

\begin{table}[ht]
\caption{Proportion of cliques that are not completely contained in at least one community -- i.e. are `split' -- by the OSLOM and MOSES overlapping CFAs. Some networks present in previous benchmarks are not shown, due to the algorithms taking too long to run.}
\label{damageDoneOverlapping}
\begin{center}
\begin{tabular}{r|p{1.5cm}|p{1.5cm}|p{1.5cm}|p{1.5cm}}
\hline
Network & OSLOM\par $>$10\% split& OSLOM \par Size $>$8 \par $>$20\% split &  MOSES \par $>$10\% split & MOSES \par Size $>$8\par $>$20\% split\\
\hline
\hline
Email-Enron  & 0.96 & 0.98 & 0.16 & 0.01 \\
Email-EuAll  & 0.93 & 0.00 & 0.06 & 0.00 \\ 
Mobile1  & 0.99 & 0.00 &0.00 & 0.00\\ %
Mobile2 & 0.94 & 0.00 &0.03 & 0.00 \\  %
Mobile3  & 0.99 & 0.00 & 0.03 & 0.00 \\ %
    \hline
Facebook-caltech  & 0.76 & 0.37 &0.41 & 0.04 \\
Facebook-princeton  & 0.93 & 0.76 &0.21 & 0.01\\ 
Facebook-georgetown  & 0.95 & 0.82 &0.22 & 0.01 \\ 
Twitter1  & 0.97 & 0.94 & 0.20 & 0.00 \\  %
Twitter2  & 0.98 & 0.91 & 0.02 & 0.00  \\  %
Twitter3 & 0.98 & 0.91 & 0.02 & 0.00 \\  %
    \hline
Collab-AstroPhysics  & 0.86 & 0.79 & 0.35 & 0.04\\ 
Collab-CondensedMatter  & 0.40 & 0.22 & 0.08 & 0.02  \\ 
Collab-HighEnergy  & 0.33 & 0.00 & 0.08 & 0.00 \\
Cite-HighEnergy  & 0.81 & 0.52 & 0.15 & 0.00 \\ 
    \hline
Amazon0302& 0.09 & 0.00  & 0.01 & 0.00  \\ 
Epinions  & 0.96 & 0.83 & 0.01& 0.00  \\ 
Wiki-Vote  & 0.79 & 0.67 & 0.04 & 0.00\\ 
    \hline
Protein-Collins & 0.13 & 0.06 & 0.06 & 0.01  \\
\hline
\end{tabular}
\end{center}
\vspace{-14px}
\end{table}

\subsection{Community Overlap Graphs}
We have considered, in Section \ref{fundamentalSection}, the fundamental partitionability of networks, by examining the connected components which exist when we only consider subsets of the networks connected by cliques.
We have also considered whether a hypergraph partitioning method, attempting to split as few cliques as possible, can partition the network, and have seen that -- assuming cliques at the core of communities -- communities overlap each other pervasively.
We can develop our intuition about these results further, by considering the possibilities for the structure present in the \emph{`Community Overlap Graph'} (COG); the graph formed by representing each community as a node, and connecting a pair of nodes (communities) by an edge, when the two communities overlap by more than some threshold number of nodes i.e. when they share more than some threshold number of nodes in common.
When considering individual cliques as our communities, this idea is identical to that of the `clique graph' discussed by Everett and Borgatti \cite{everett1998analyzing} and discussed further by Evans \cite{evans2010clique}.
\begin{figure}[!htb]
\begin{center}
    \subfigure[]{
        \includegraphics[width=52mm]{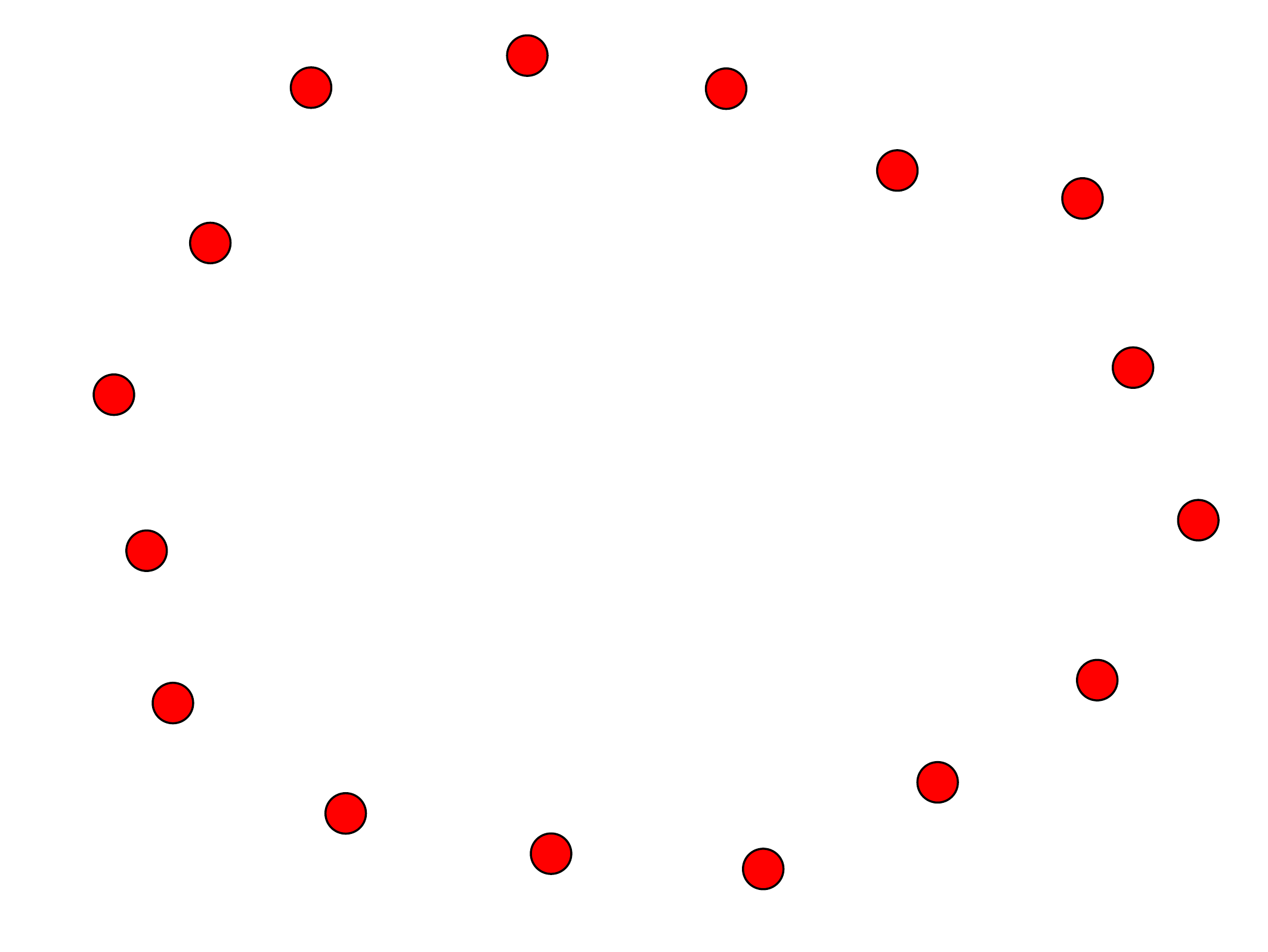}
        \label{1}
    }
    \hspace{32px}
    \subfigure[]{
        \includegraphics[width=52mm]{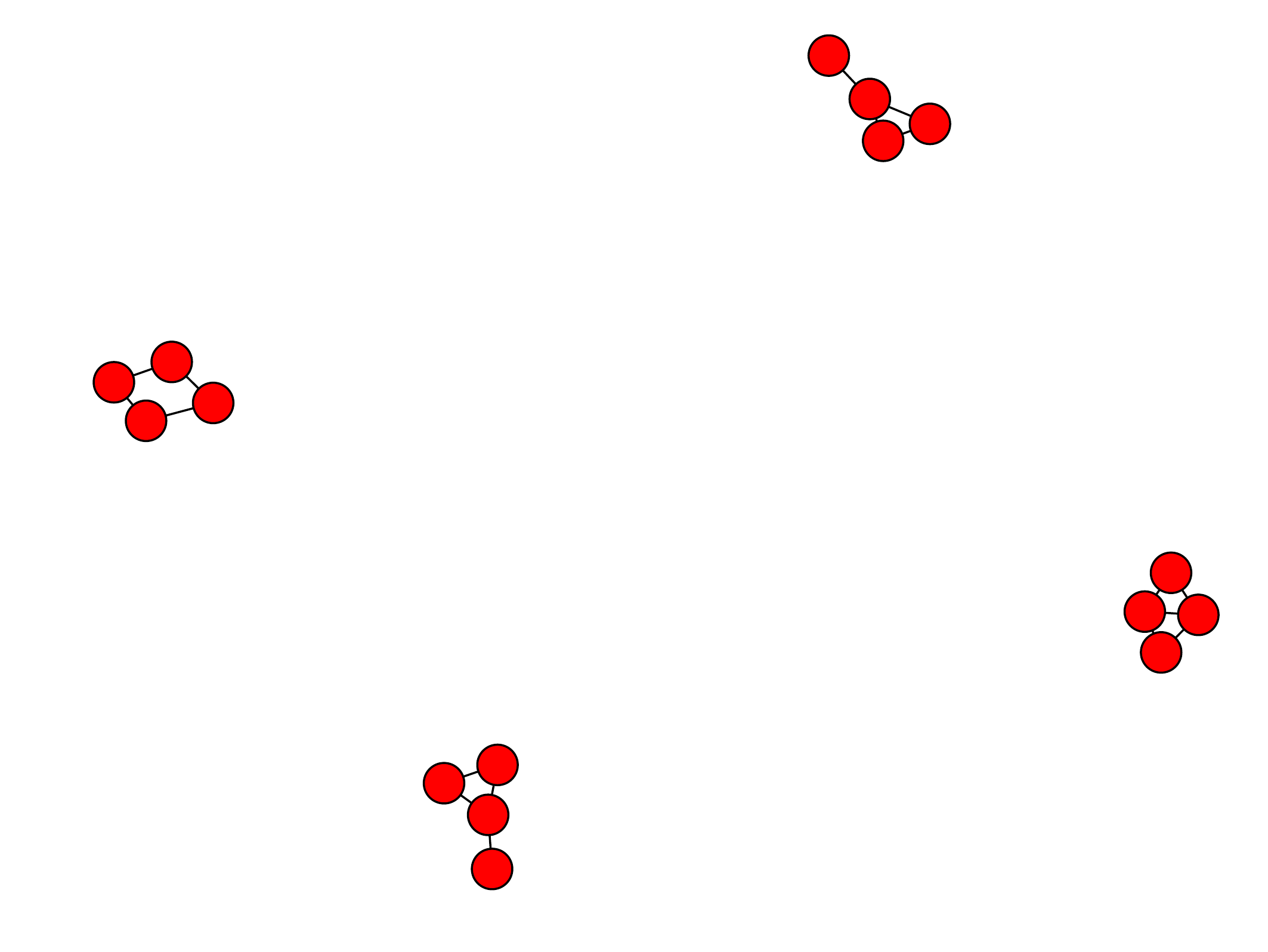}
        \label{2}
    }
    \subfigure[]{
        \includegraphics[width=52mm]{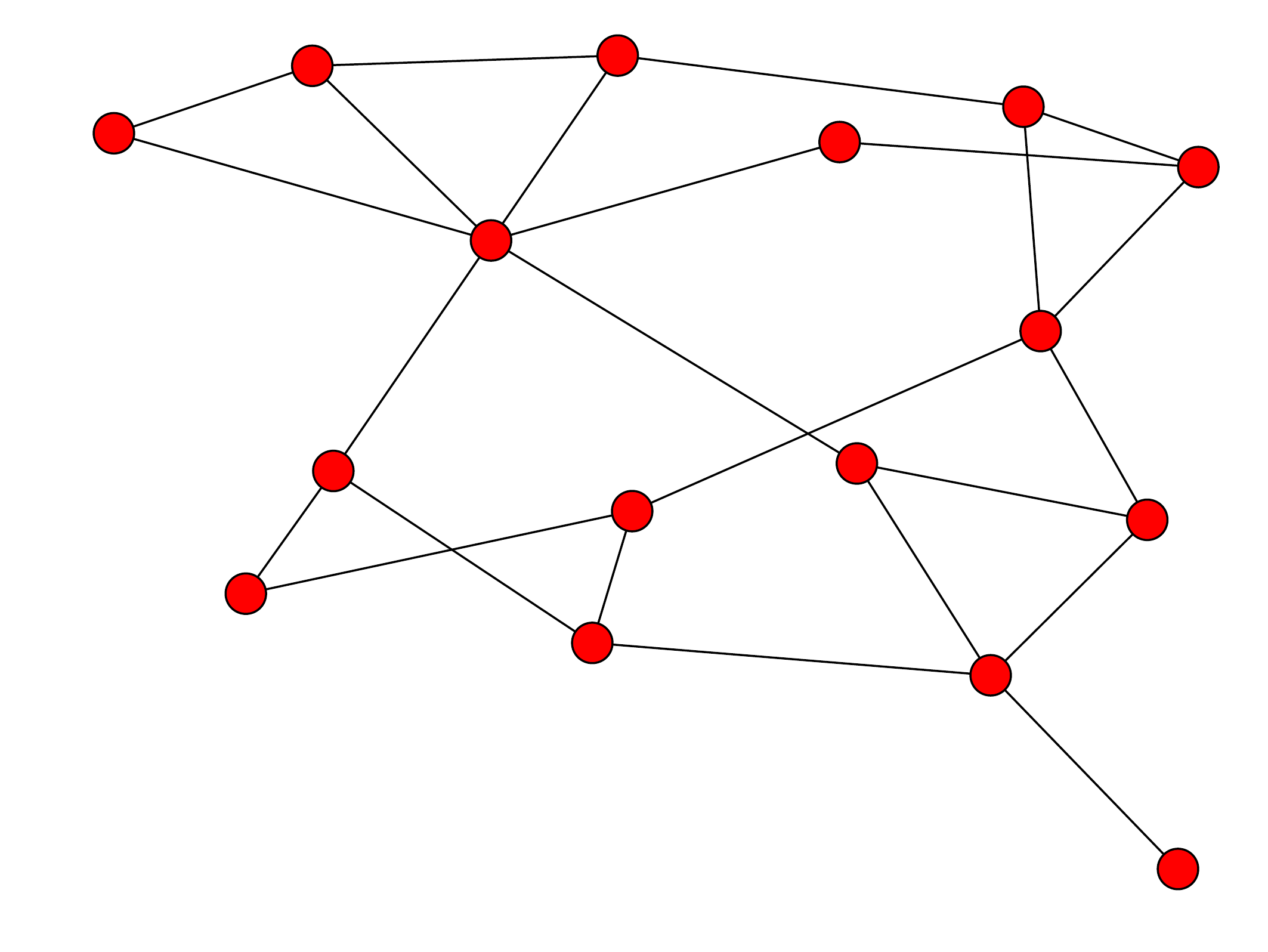}
        \label{3}
    }
\end{center}
\caption{Illustration of some possibilities for the Community Overlap Graph, for a network with 16 communities. Each node represents a community; edges connect pairs of overlapping communities. (a) Non-overlapping communities. (b) Overlapping communities, but clustered, with no path through overlap. (c) Overlapping communities with unpartitionable overlap.}
\label{communityOverlapGraph}
\end{figure}

However, the large numbers of maximal cliques present in the networks we study make explicitly working with the clique graph difficult.
The communities found by overlapping CFAs however, are typically smaller in number (as shown in Tables \ref{cogMOSES} and \ref{cogOSLOM}).
Different possibilities for what structure we might see in the community overlap graph are shown in Figure \ref{communityOverlapGraph}.
We can see that Figure \ref{communityOverlapGraph}(a) corresponds to a world view of non overlapping communities, in which the partitioning of networks into communities makes obvious sense.
Figure \ref{communityOverlapGraph}(b) contains overlapping communities, but, perhaps surprisingly, it still makes some sense to partition the network, with partitions dividing clusters of overlapping communities together.
We have shown, from our analysis of paths through cliques, and attempting to partition the network using hypergraph partitioning on the found cliques, that a world-view similar to Figure \ref{communityOverlapGraph}(c) is most appropriate.%
We will now discuss these ideas in more detail, with reference to actual community overlap graphs, generated from the communities found by the two overlapping community finding algorithms, on empirical data.

\subsection{Analysis of Community Overlap Graphs of Overlapping CFAs}
\begin{figure}[!hp]
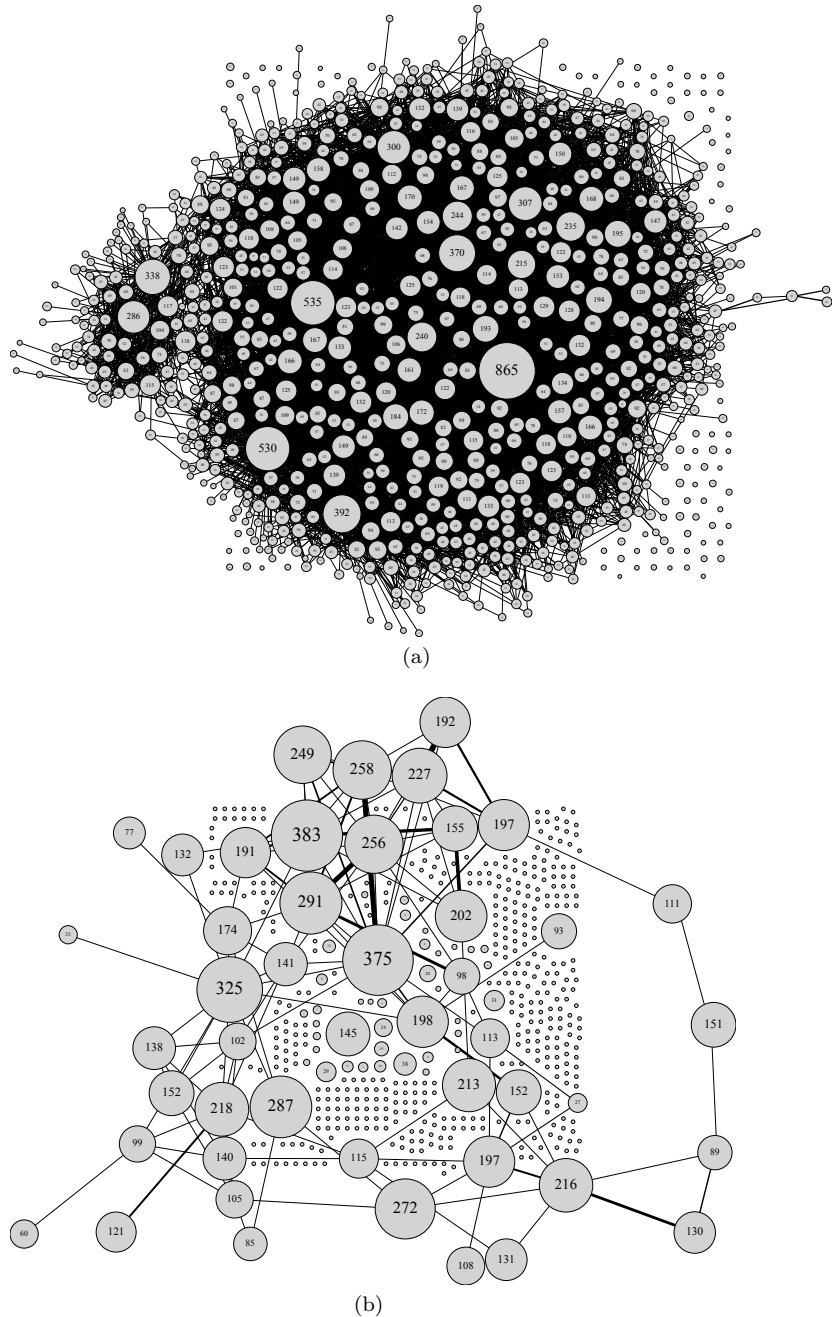

    \subfigure[]{
        \includegraphics[width=4.3in]{Facebook_princeton_moses_overlap4_graphviz_dot.pdf}
        \label{empiricalCOG_princeton_moses}
    }
    \subfigure[]{
\includegraphics[width=3.8in]{Facebook_princeton_oslom_overlap4_graphviz_dot.pdf}
\label{empiricalCOG_princeton_oslom}
    }
\caption{Visualisation of COG of Facebook Princeton network. An edge is drawn whenever two communities overlap by at least 4 nodes. Edge width is proportional to overlap, and node area is proportional community size. Communities are labeled with the number of nodes they contain. Nodes may be present in multiple communities; two communities with a high degree of overlap contain fewer unique nodes than the sum of their labels. Shown here is the COG extracted from running (a) MOSES and (b) OSLOM on the Facebook Princeton network. Networks visualised with Graphviz \cite{ellson2004graphviz} force directed layout.}
\label{empiricalCOGs_princeton}
\end{figure}

In Figure \ref{empiricalCOGs_princeton} we show visualisations of the Community Overlap Graph of the communities found by the MOSES and OSLOM algorithms, in the \mbox{Facebook} Princeton network.
In order to show only the more significant overlaps, we draw an edge between two communities overlapping by at least 4 nodes.
These visualisations show that in this particular network, most of the larger communities found by these two algorithms -- and hence most of the nodes in the network -- are part of a connected component of overlapping communities. 
As such, the empirical visualization corresponds most closely to Figure \ref{communityOverlapGraph}(c), and hence partitioning to find communities is not suitable in networks like this.
Visualizing the community overlap graphs of these networks shows clearly the extent to which communities overlap, and the structure that would be broken by partitioning these networks in order to find communities.

In addition to visualizing these networks, we can attempt to quantify the degree to which a set of overlapping communities is partitionable, similar to how we examined the fundamental partitionability of networks in Section \ref{fundamentalSection}.
We attempt to quantify this by examining how many of the communities in the community overlap graph are in the largest connected component of that graph, and what proportion of the nodes in the source network they contain.
If a large proportion of communities and nodes are in a connected component, then this again would indicate quantitatively that the community structure is closer to Figure \ref{communityOverlapGraph}(c) than (a) or (b).

Our results for the MOSES method are presented in Table \ref{cogMOSES}, and for OSLOM in Table \ref{cogOSLOM}.
As we can see, in line with our earlier results using cliques, the degree of overlap varies across networks with the social networks -- particularly the Facebook networks -- showing the least partitionable results.

In line with the results obtained by quantifying the proportion of cliques split, MOSES finds structure that is more highly overlapping than OSLOM.
These results show an interesting method of quantifying the degree of overlap of community structure in a given network, and for a given overlapping CFA.

\begin{figure}[!htb]
\centering
\includegraphics[width=3.2in]{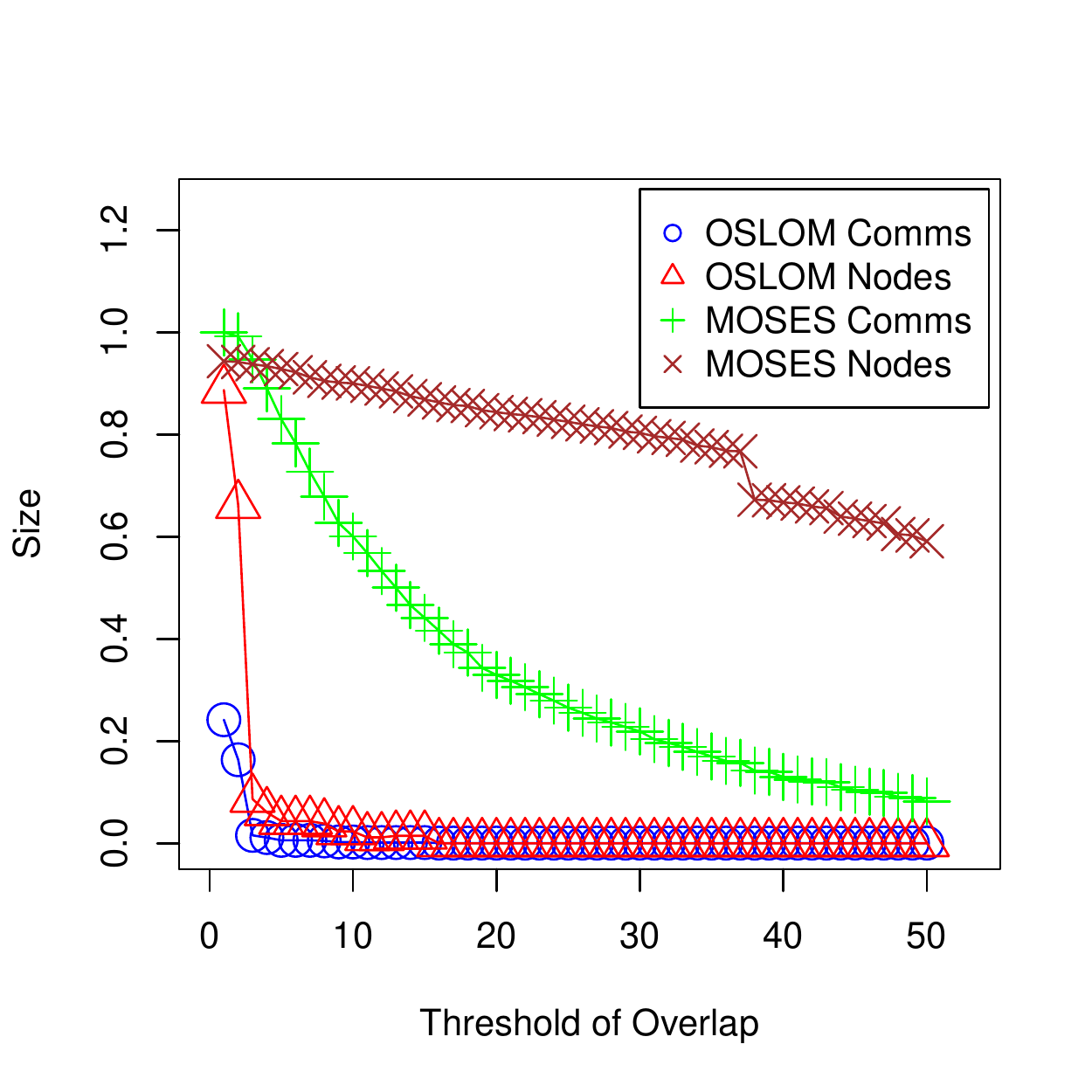}
\caption{As the threshold of overlap is changed, the size of the largest connected component in the community overlap graph changes.  We investigate how the size of this component varies both in terms of the number of communities in it, and the total number of nodes connected by communities that overlap by at least that threshold. We display the size of the largest component, both in terms of the proportion of communities that are in it, and in terms of the proportion of nodes in the underlying graph which are in it, for both OSLOM and MOSES Overlapping CFAs}
\label{COGThreshold}
\end{figure}

\begin{table}[!ht]
\begin{center}
\caption{Results for the size of the largest connected component (LCC) of the \mbox{community} overlap graph (COG), examining community structure found by MOSES.}
\label{cogMOSES}
\begin{tabular}{l|l|p{1.5cm}|p{1.5cm}|p{1.5cm}|p{1.8cm}}
  \hline
Network & Nodes & Number Comm-unities & Comms in LCC of COG & Nodes in LCC of COG & Proportion Nodes in LCC COG \\ 
    \hline
    \hline
            Email-Enron & \numprint{36692} & \numprint{2471} & \numprint{587} & \numprint{14573} & 0.4\\ 
            Email-EuAll & \numprint{265009} & \numprint{473} & \numprint{257} & \numprint{24919} & 0.09\\ 
                 Mobile1 & \numprint{10001} & \numprint{437} & \numprint{38} & \numprint{1159} & 0.12\\ 
                 Mobile2 & \numprint{10001} & \numprint{323} & \numprint{219} & \numprint{8609} & 0.86\\ 
                 Mobile3 & \numprint{10001} & \numprint{171} & \numprint{120} & \numprint{8478} & 0.85\\ 
    \hline
        Facebook-caltech & \numprint{769} & \numprint{81} & \numprint{71} & \numprint{666} & 0.87\\ 
     Facebook-princeton & \numprint{6596} & \numprint{797} & \numprint{710} & \numprint{6162} & 0.93\\ 
   Facebook-georgetown & \numprint{9414} & \numprint{893} & \numprint{800} & \numprint{8740} & 0.93\\ 
                Twitter1 & \numprint{2001} & \numprint{161} & \numprint{132} & \numprint{1686} & 0.84\\ 
                Twitter2 & \numprint{2001} & \numprint{129} & \numprint{99} & \numprint{1680} & 0.84\\ 
                Twitter3 & \numprint{2001} & \numprint{188} & \numprint{101} & \numprint{1080} & 0.54\\ 
    \hline
     Collab-AstroPhysics & \numprint{18771} & \numprint{2816} & \numprint{677} & \numprint{9953} & 0.53\\ 
Collab-CondMat & \numprint{23133} & \numprint{3458} & \numprint{175} & \numprint{3760} & 0.16\\ 
      Collab-HighEnergy & \numprint{9875} & \numprint{1663} & \numprint{15} & \numprint{274} & 0.03\\ 
        Cite-HighEnergy & \numprint{27769} & \numprint{1625} & \numprint{998} & \numprint{21445} & 0.77\\ 
    \hline
            Amazon0302 & \numprint{262111} & \numprint{23665} & \numprint{47} & \numprint{1767} & 0.01\\ 
                Epinions & \numprint{75879} & \numprint{795} & \numprint{249} & \numprint{19889} & 0.26\\ 
              Wiki-Vote & \numprint{7115} & \numprint{65} & \numprint{63} & \numprint{3805} & 0.53\\ 
    \hline
          Protein-Collins & \numprint{1622} & \numprint{150} & \numprint{16} & \numprint{221} & 0.14\\ 
    \hline
    \end{tabular}
    \end{center}
\vspace{2px}
    \end{table}
\begin{table}[!ht]
\begin{center}
\caption{Results for the size of the largest connected component (LCC) of the \mbox{community} overlap graph (COG), examining community structure found by OSLOM.}
\label{cogOSLOM}
\begin{tabular}{l|l|p{1.5cm}|p{1.5cm}|p{1.5cm}|p{1.8cm}}
  \hline
Network & Nodes & Number Comm-unities & Comms in LCC of COG & Nodes in LCC of COG & Proportion Nodes in LCC COG \\ 
    \hline
    \hline

            Email-Enron & \numprint{36692} & \numprint{10620} & 46 & \numprint{2722} & 0.07\\ 
            Email-EuAll & \numprint{265009} & \numprint{131143} & - & - & -\\ 
                 Mobile1 & \numprint{10001} & \numprint{8435} & 1 & 3 & 0\\ 
                 Mobile2 & \numprint{10001} & \numprint{8119} & 4 & 235 & 0.02\\ 
                 Mobile3 & \numprint{10001} & \numprint{9195} & 2 & 157 & 0.02\\ 
    \hline
        Facebook-caltech & \numprint{769} & \numprint{137} & 2 & 100 & 0.13\\ 
     Facebook-princeton & \numprint{6596} & \numprint{920} & 11 & 404 & 0.06\\ 
   Facebook-georgetown & \numprint{9414} & \numprint{1189} & 5 & 178 & 0.02\\ 
                Twitter1 & \numprint{2001} & 467 & 21 & \numprint{1529} & 0.76\\ 
                Twitter2 & \numprint{2001} & 113 & 7 & \numprint{1463} & 0.73\\ 
                Twitter3 & \numprint{2001} & 289 & 9 & \numprint{1040} & 0.52\\ 
    \hline
     Collab-AstroPhysics & \numprint{18771} & \numprint{4106} & 9 & 202 & 0.01\\ 
Collab-CondMat & \numprint{23133} & \numprint{5911} & 6 & 159 & 0.01\\ 
      Collab-HighEnergy & \numprint{9875} & \numprint{3808} & 7 & \numprint{145} & 0.01\\ 
        Cite-HighEnergy & \numprint{27769} & \numprint{4393} & 12 & \numprint{358} & 0.01\\ 
    \hline
            Amazon0302 & \numprint{262111} & \numprint{37374} & 19 & \numprint{424} & 0\\ 
                Epinions & \numprint{75879} & \numprint{46260} & 57 & \numprint{3739} & 0.05\\ 
              Wiki-Vote & \numprint{7115} & \numprint{2744} & 21 & \numprint{4085} & 0.57\\ 
    \hline
          Protein-Collins & \numprint{1622} & 529 & 2 & 50 & 0.03\\ 

    \hline
    \end{tabular}
    \end{center}
\vspace{-16px}
    \end{table}

In Tables \ref{cogMOSES} and \ref{cogOSLOM} we used an overlap of 4 nodes as a threshold for `significant' overlap between two communities.
It is interesting to examine how the threshold used to analyze the community overlap graph effects the connectivity of that graph.
Figure \ref{COGThreshold} shows how the threshold effects the size of the largest connected component in the community overlap graph, both in terms of the number of communities in it, and the number of nodes of the underlying network that are in it, for both of the overlapping CFAs we examined. 
It can be seen from this figure that OSLOM has a sharp falloff in the size of the LCC as the threshold of overlap is increased.
MOSES exhibits a much more gradual falloff in the size of the LCC -- even if we require that communities have 10 nodes in common for them to be overlapping, the graph is still largely unpartitionable, without breaking several community overlaps.
This difference between MOSES and OSLOM is perhaps not surprising, given MOSES's tendency to find larger communities than OSLOM, and is consistent with the results in terms of the proportion of cliques split.
This is further evidence that the level of overlap in the structures found by varying overlapping CFAs can vary widely.

\section{Conclusion}
We have investigated a wide range of empirical networks, characterising them according to the proportion of cliques in them that are split by various partitioning methods.
Our results show that the early intuition on how communities are embedded in graphs does not hold across all networks and domains.
On many complex networks cliques do not exist \emph{solely} in community cores connected only by narrow bridges and weak ties -- instead they frequently overlap across the community boundaries produced by partitioning algorithms.

If we accept cliques as conservative lower bounds for community structure, then, on many networks, partitioning CFAs are fundamentally limited in the completeness of the communities they can find, as shown by our results on the graph of edges in cliques, and from using hypergraph partitioning algorithms to partition cliques.
This shows that communities are not easily separable from each other simply by removing structurally weak ties; instead, communities overlap across each other, with pairs of community frequently connected by strong ties, and other communities.

Our analysis of overlapping community finding algorithms has shown that some overlapping CFAs produce sets of communities in which each individual clique is fully contained.
However, as we have shown, not all overlapping CFAs satisfy this property.
We have presented community overlap graphs as another tool -- in addition to cliques -- with which to explore the output of overlapping CFAs. 
We have shown that on some networks, the communities as output by overlapping CFAs reveal a community structure that cannot be partitioned.

Thus, caution is warranted when using partitioning community finding algorithms where there is a sensitivity requirement that all significant community structure be found.
In agreement with recent research on pervasive overlap, conceptualising networks as overlapping meshes of strong ties, with denser community regions, and using a CFA designed to find communities that overlap, will be more appropriate in many application domains.

\section{Further Work}
Work on formal models of community generation that might explain whether a network is suitable for partitioning, and attempt to characterise the generative processes behind this global overlap would be interesting.
That cliques frequently span communities also has implications for the type of diffusion processes that can occur on networks; data on the non-partitionable nature of communities may lead to an enhanced understanding of diffusion on complex networks.
We hope that studying the nature of community overlap can lead to a better fundamental understanding of structure in empirical networks, and help development of future community finding algorithms.

\section{Acknowledgements}
This work is supported by Science Foundation Ireland under grant 08/SRC/I1407: Clique: Graph and Network Analysis Cluster.
An earlier version of this work appeared in \emph{ASONAM '11} \cite{reid2011partitioning}.

\bibliographystyle{abbrv}
\bibliography{mybib}

\end{document}